\documentclass[12pt,twocolumn]{emulateapj}

\tightenlines

\usepackage{ulem} 
\usepackage{xcolor}
\usepackage{graphicx}
\usepackage{enumerate}

\newcount\listnorom
\newcommand\listromanDE{\global\advance \listnorom by 1
{\lowercase\expandafter{(\romannumeral\listnorom)}\ }}

\newcount\listnumber
\newcommand\listDE{\global\advance \listnumber by 1
{\lowercase\expandafter{(\number\listnumber)}\ }}

\def\lsim{\raise0.3ex
  \hbox{$<$\kern-0.75em\raise-1.1ex\hbox{$\sim$}}\,}
\def\gsim{\raise0.3ex
  \hbox{$>$\kern-0.75em\raise-1.1ex\hbox{$\sim$}}\,}

\newcommand{\AstroH}{\textit{Astro-H}}

\newcommand{\Zave}{\langle Z \rangle}

\newcommand{\fDamp}{f_\mathrm{damp}}

\newcommand{\crhydro}{{\it CR-hydro-NEI}}

\newcommand{\SunMyr}{$\Msun$\,yr$^{-1}$}

\newcommand{\Rej}{\rho_\mathrm{ej}}

\newcommand{\Rwind}{\rho_\mathrm{wind}}
\newcommand{\Vwind}{V_\mathrm{wind}}

\newcommand{\pcc}{cm$^{-3}$}
\newcommand\Msun{M_{\odot}}

\newcommand{\Teq}{t_\mathrm{eq}}

\newcommand{\EffDSA}{{\cal E_\mathrm{DSA}}}
\newcommand{\EffDSAfs}{{\cal E_\mathrm{DSA}^\mathrm{FS}}}
\newcommand{\EffDSArs}{{\cal E_\mathrm{DSA}^\mathrm{RS}}}

\newcommand{\Kep}{K_\mathrm{ep}}

\newcommand{\SA}{semi-analytic}
\newcommand{\NT}{non-thermal}

\newcommand{\DSA}{diffusive shock acceleration}

\newcommand{\MFA}{magnetic field amplification}

\newcommand{\kmps}{km s$^{-1}$}
\newcommand{\NL}{nonlinear}
\newcommand{\gamray}{$\gamma$-ray}

\newcommand{\Alf}{Alfv\'en}

\newcommand{\SNRJ}{SNR RX J1713.7-3946}

\newcommand{\SC}{self-consistent}
\newcommand{\SCly}{self-consistently}

\newcommand{\muG}{$\mu$G}
\newcommand{\nISM}{n_\mathrm{ISM}}

\newcommand{\be}{\begin{eqnarray}}
\newcommand{\ee}{\end{eqnarray}}

\newcommand{\rel}{relativistic}

\newcommand{\mc}{Monte Carlo}

\newcommand{\mrm}{\mathrm}

\newcount\Inum
\newcount\IInum
\newcount\IIInum
\newcount\IVnum
\def\I{\global\multiply\IInum by 0 \global\multiply\IIInum by 0
            \global\multiply\IVnum by 0 \global\advance \Inum by 1
            {\the\Inum. }}
\def\II{\global\multiply\IIInum by 0\global\multiply\IVnum by 0
       \global\advance \IInum by 1 {\the\Inum.\the\IInum. }}
\def\III{\global\multiply\IVnum by 0\global\advance \IIInum by 1
            {\the\Inum.\the\IInum.\the\IIInum. }}
\def\IV{\global\advance \IVnum by 1
            {\the\IVnum. }}

\begin{document}

\title{Reverse and forward shock X-ray emission in an
evolutionary model of supernova remnants 
undergoing efficient diffusive shock acceleration}

\author{
Shiu-Hang Lee\altaffilmark{1,2},
Daniel J. Patnaude \altaffilmark{3},
Donald C. Ellison \altaffilmark{4},
Shigehiro Nagataki \altaffilmark{2} and
Patrick O. Slane \altaffilmark{3} 
}

\altaffiltext{1}{Institute of Space and Astronautical Science, Japan Aerospace Exploration Agency,
3-1-1 Yoshinodai, Chuo-ku, Sagamihara, Kanagawa 252-5210, Japan; 
slee@astro.isas.jaxa.jp}
\altaffiltext{2}{RIKEN, Astrophysical Big Bang Laboratory, 
2-1 Hirosawa, Wako, Saitama 351-0198, Japan; 
shiu-hang.lee@riken.jp; shigehiro.nagataki@riken.jp}
\altaffiltext{3}{Harvard-Smithsonian Center for Astrophysics, 
60 Garden Street, Cambridge, MA 02138, USA;
slane@cfa.harvard.edu; dpatnaude@cfa.harvard.edu}
\altaffiltext{4}{Physics Department, North Carolina State
University, Box 8202, Raleigh, NC 27695, USA;
don\_ellison@ncsu.edu}

\begin{abstract}
We present new models for the forward and reverse shock thermal X-ray emission from core-collapse and Type Ia supernova remnants (SNRs) which include the efficient production of cosmic rays (CR) via nonlinear diffusive shock acceleration (DSA). Our \crhydro\ code takes into account non-equilibrium ionization, hydrodynamic effects of efficient CR production on the SNR evolution, and collisional temperature equilibration among heavy ions and electrons in both the shocked supernova (SN) ejecta and the shocked circumstellar material.  While X-ray emission is emphasized here, our code \SCly\ determines both thermal and non-thermal broadband emission from radio to TeV energies. We include Doppler broadening of the spectral lines by thermal motions of the ions and by the remnant expansion. We study, in general terms, the roles which the ambient environment, progenitor models, temperature equilibration, and processes related to DSA have on the thermal and non-thermal spectra. The study of X-ray line emission from young SNRs is a powerful tool for determining specific SN elemental contributions,  and for providing critical information that helps to understand the type and energetics of the explosion, the composition of the ambient medium in which the SN exploded, and the ionization and dynamics of the hot plasma in the shocked SN ejecta and interstellar medium. With the approaching launch of the next-generation X-ray satellite \textit{Astro-H}, observations of spectral lines with unprecedented high resolution will become a reality. Our \SC\ calculations of the X-ray spectra from various progenitors will help interpret future observations of SNRs.
\end{abstract}


\section{Introduction}
While some heavy elements are synthesized during the stable nuclear burning that occurs during the normal evolutionary sequence for a star, the bulk of the elements with nuclear charge $>$ 8 are produced during either the brief  destabilization of the core of a massive star which leads to a core-collapse supernovae (CCSNe), or the thermonuclear detonation of a Chandrasekhar mass white dwarf. These two types of events are thought to produce the  bulk of the high $Z$ elements such as silicon and iron. 

When the progenitor explodes, these elements are ejected and a strong
forward shock (FS) is driven out into the surrounding circumstellar medium (CSM), while another shock, the reverse shock (RS), is driven back into the expanding bubble of ejecta, which is heated to millions of degrees Kelvin, and ionized to H- and He-like charge states. 
The thermal X-ray emission reveals details of the explosion type (core collapse (CC) versus thermonuclear) and energy, and in some cases can point to a possible explosion mechanism. It also encapsulates the evolutionary history of the shock-heated plasma, and the X-ray emission at the FS gives a direct measure of the composition of the interstellar medium (ISM).

In addition to heating, the collisionless shocks in a supernova remnant (SNR) also accelerate some fraction of the ISM, and perhaps the ejecta material, to \rel\ energies via \DSA\ (DSA). This cosmic-ray (CR) acceleration process has been studied extensively for 
thermal and non-thermal emission from the FS in evolving SNRs using the \crhydro\ code we employ here 
\citep[e.g.,][]{EPSBG2007,PES2009,PSRE2010,EPSR2010,CSEP2012,LSENP2013}. 
A key result from this work is that the modification of the pressure and density by efficient DSA needs to be taken into account when computing the accompanying thermal spectrum: CR production and thermal X-ray emission are not independent of each other.
In this paper, we present initial results from the \crhydro\ code that has been generalized to include thermal emission from the RS simultaneously with broad-band emission from the FS. 
We present only generic results here and leave the precise fitting of individual SNRs to future work.

There have been numerous previous studies of X-ray emission from shocked supernova (SN) ejecta for both young and middle
aged SNRs \citep[e.g.,][]{HSC83,hughes85,deb94,nymark06}. 
However, in general, the thermal X-ray emission has not been calculated self-consistently in many hydrodynamical models, including some that consider nonlinear DSA. Exceptional examples include: \citet{PES2009, PSRE2010}, but the RS has not been considered in these studies; three-dimensional simulations by \citet{ferrand12}, but they have not considered the spatial distribution of chemical elements in the ejecta; and also \citet{badenes03} (and subsequent papers) which have treated the ejecta in detail, but have not included DSA and computed non-equilibrium ionization (NEI) only as a post-process with a fixed post-shock electron-to-proton temperature ratio.
While the X-ray emission does not represent a significant cooling element in non-radiative SNRs, and thus does not affect the hydrodynamics, it provides an observational diagnostic that can be crucial in interpreting the nature of gamma-ray emission produced in the DSA process. This has been shown in some recent studies of such emission from DSA at the FS \citep[e.g.,][]{EPSR2010,CSEP2012,LSENP2013}. 
The model we present here is the first to describe the 
broadband emission (thermal and non-thermal) of a young, shell-type SNR from both the shocked CSM and {\it shocked ejecta} using a self-consistent calculation of the hydrodynamics, collisional ionization, ion/electron temperature equilibration, and non-linear particle acceleration. 
We treat elemental compositions derived from both CC and thermonuclear SN explosion models. To accommodate arbitrary chemical compositions, including those found in SN ejecta, we have generalized our temperature equilibration routine to follow the heating of a number of elements and include 140 ionic states in all. 

In addition, we now include thermal and Doppler broadening
effects on the emitted X-ray spectrum. The thermal 
broadening allows us to estimate the ion temperature which can be directly compared
against future high spectral resolution X-ray observations of SNR ejecta such as those expected from, for example,   
{\it Astro-H}. 
In principle, detailed observations of ion temperatures will allow us to determine if the collisionless heating is substantially faster than that expected from Coulomb equilibration alone, as suggested by recent studies of ejecta in Tycho's SNR \citep{Yamaguchi2014}.

In Section~\ref{model}, we describe the newly implemented features of our code and give parameters 
that are used for our hydrodynamical calculations for different SN ejecta models. 
In Section~\ref{result}, we present results for both young Type Ia and 
CCSN remnants, paying particular attention to the thermal X-ray emission from the shocked ejecta and ambient medium. 
We also highlight some of the physical parameters which influence the thermal X-ray emission 
(e.g., the shape of the continuum, line strengths, and line centroids) and discuss their 
implications for interpreting future X-ray observations of SNRs.
Finally, we summarize our results in Section~\ref{conclusion}, and discuss future applications of our code with a goal toward better understanding the population of Galactic and nearby extragalactic SNRs, as well as addressing questions on the origin of Galactic CRs.

\section{Model}
\label{model}

In this section we describe the modifications to the \crhydro\ code that allow us to simultaneously describe thermal and non-thermal emission at the FS and RS of young SNRs.
  
\subsection{Ejecta Models}

\begin{figure}
\centering
\includegraphics[width=0.45\textwidth]{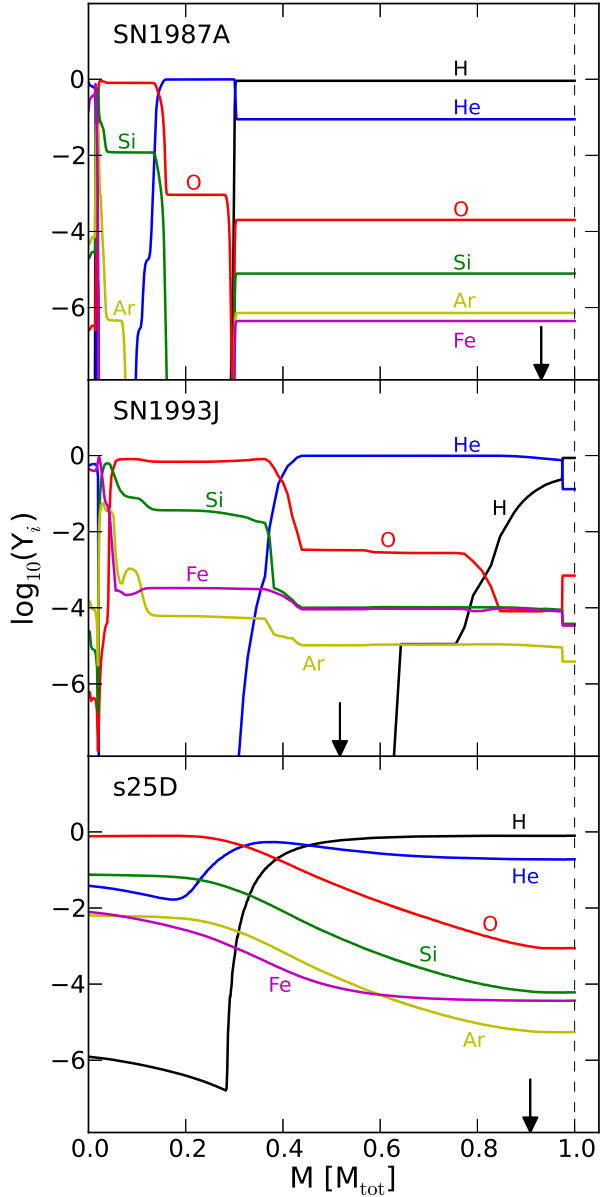}
\caption{
Molar fractional abundances as a function of enclosed mass (minus a central neutron star and normalized to the total ejecta mass) for H, He, O, Si, Ar and
Fe, for CCSN models for SN1987A (upper panel), SN1993J (middle panel) and 
s25D (lower panel). The vertical dashed lines indicate the outer surface of each ejecta model. The arrows indicate the final locations of the RS at the end of our simulations at 500~yr. (See text and Table~\ref{table:ejecta} for more details of these ejecta models)
}
\label{fig:ccsne}
\end{figure}

For this initial study, we make use of published results for the composition of the ejecta in
both core-collapse (CC) and Type Ia SNe. The models are summarized in Table~\ref{table:ejecta}. 
For the CCSNe, we make use of models developed initially by Heger \& Woosley 
\citep[e.g.,][]{rauscher02,woosley02,joggerst09}, and in particular make use of model s25D \citep{Heger2010}, which
corresponds to a 25~$M_\sun$ main sequence progenitor with initially solar metallicity. 
We chose this model mainly because it has a metallicity appropriate for Galactic main sequence stars (as opposed
to say, zero metallicity progenitors or progenitors of SNe in the Magellanic clouds where the
initial metallicity is $\approx$ 0.3~$Z_\sun$). For the s25D model, the stellar evolution is followed
and the progenitor loses $\approx$ 8 $M_\sun$ of material to wind loss prior to the onset of the
core-collapse. Details of the explosion can be found in e.g. \citet{joggerst09}, but to summarize,
a piston is placed at the base of the oxygen burning layer. The piston has a mass of a typical neutron
star and energy of $1.2\times10^{51}$ erg. The KEPLER code is used to simulate the stellar evolution
and explosive nucleosynthesis, and uses a nuclear reaction network of 19 isotopes to calculate 
the final isotopic yield. The resultant abundances as a function of enclosed mass for model s25D
are shown in the bottom panel of Figure~\ref{fig:ccsne}. For this model, a prescription for mixing
between mass layers has been applied in an ad hoc manner \citep{woosley02}.

We also include a model for CCSN ejecta described in \citet{Saio1988a, Saio1988b}, \citet{Hashimoto1989} and \citet{Shigeyama1990}.
This model has been tailored to match observations of SN1987A in the Large Magellanic Cloud.
The model comprises a 6 $M_\sun$ He star enclosed by a $\sim$ 10 $M_\sun$ hydrogen envelope.
Finally, we include a model that has previously been applied to the Type IIb SN~1993J \citep{Shigeyama1994}.
The progenitor for this model consists of a zero-age main sequence 18 $M_\sun$ star with 
metallicity $Z = 0.02$ $Z_\sun$. The progenitor loses 15 $M_\sun$ of material prior to core-collapse. A
thin layer of hydrogen with mass 0.08 $M_\sun$ encloses the interior of the ejecta, post-SN.
This particular model has been applied by \citet{Nozawa2010} to the young CC SNR Cassiopeia A (Cas A).
While our goal here is not to model any particular SNR, for simplicity we refer to these models as ``SN1987A'' and ``SN1993J''.
\footnote{We note that no mixing by the Rayleigh-Taylor instability and fallback is applied in these two ejecta models.}

\begin{figure}
\centering
\includegraphics[width=0.45\textwidth]{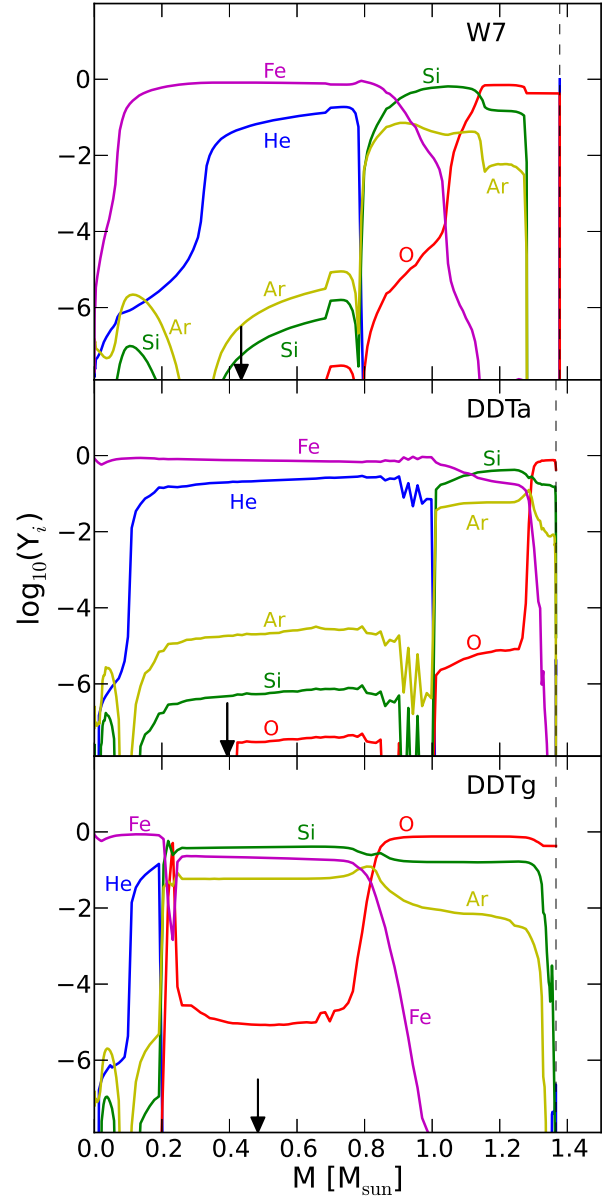}
\caption{
Molar fractional abundances as a function of enclosed mass for He, O, Si, Ar and
Fe for Type Ia SN models W7 (upper label), DDTa (middle panel) and DDTg (lower panel).
The vertical dashed lines and arrows have the same meanings as in Figure~\ref{fig:ccsne}.
}
\label{fig:type1a}
\end{figure}

The Type Ia models are obtained from the detonation of a 1.38 $M_{\sun}$ 
carbon+oxygen white dwarf in hydrostatic equilibrium. The detonation
propagation was determined by self consistently solving the hydrodynamical
and nuclear evolutionary equations. For this study, we chose two delayed-detonation
models: DDTa, and DDTg \citep{Badenes2008}, shown in Figure~\ref{fig:type1a}. Model DDTa
represents an energetic ($E_{\mathrm{SN}}$ = 1.4 $\times$ 10$^{51}$ erg)
explosion producing $\sim$ 1 $M_{\sun}$ of $^{56}$Ni, while model
DDTg is a subenergetic model with $E_{\mathrm{SN}}$ = 0.9 $\times$ 10$^{51}$
erg and 0.3 $M_{\sun}$ of $^{56}$Ni. The difference in synthesized nickel
mass is reflected in the amount of Fe shown in Figure~\ref{fig:type1a}.
In addition to these detonation models, we also include W7, a widely used pure carbon-deflagration model developed by 
\citet{Nomoto1984}. 

In all cases, the simulations provide detailed information on the composition as a function of mass 
coordinate and on the density and velocity profiles for the expanding ejecta. 
For our purposes, however, we simply take the composition as a function of mass coordinate at the end of the explosion simulation and map this to either an exponential ejecta density profile for Type Ia SNe \citep[e.g.,][]{DC98} or a power-law profile (i.e., $\Rej \propto r^{-n}$) for CCSNe \citep[e.g.,][]{ch82,TM99}, with a plateau in the inner core. 
The index $n$ is a free parameter in our models chosen to best fit the density profile from the SN explosion simulations. 

The circumstellar environments are also complicated and influenced
by the detailed evolution of the progenitors \citep[see, for example,][]{Dwarkadas2005}. 
The CC models
provide some information on the mass loss history of the progenitors, while the
Type Ia models do not. For simplicity we model the circumstellar environments of our Type Ia SNe as having a constant density and magnetic field, and we assume the 
CCSNe explode in a pre-SN wind with density $\Rwind \propto r^{-2}$ and magnetic field 
determined by the assumption of magnetic flux conservation \citep[i.e., Equation~2.3 in][]{CL94}. 
Note that the total swept-up masses of CSM material by the FS in 
our CCSN models stay in the range of $1.1-1.4$~M$_\odot$, which are smaller than the total $M_\mrm{loss}$ in their pre-SN winds (see Table~\ref{table:ejecta}), therefore it is justified to ignore the transition region where the winds terminate and merge with the undisturbed ISM, in our hydro simulations.
\begin{deluxetable*}{lccccl} 
\tablecolumns{6}
\tablecaption{Summary of Ejecta Models}
\tablehead{
\colhead{Name} & \colhead{SN Type}  & \colhead{$E_\mrm{SN}$ ($10^{51}$ erg)} & \colhead{$M_\mrm{ej}$ ($M_\odot$)\tablenotemark{a}} & \colhead{$M_\mrm{loss}$ ($M_\odot$)}
& \colhead{Main Reference}
}
\startdata
DDTa 	& Delayed detonation Ia		& 1.27	& 1.38	& n/a 			& Badenes et al. (2008) \\
DDTg 	& Delayed detonation Ia		& 0.85	& 1.38	& n/a 			& Badenes et al. (2008) \\
W7 		& Deflagration Ia			& 1.23	& 1.38	& n/a				& Nomoto et al. (1984) \\
s25D 	& CCSN with ad hoc mixing	& 1.21	& 12.3	& $\approx 8$		& \citet{Heger2010} \\
SN1987A 	& II-P					& 1.11	& 14.8	& $\sim 6-8$ 		& Saio et al. (1988a) \\
SN1993J 	& IIb						& 0.98	& 2.94	& 15.1			& Nozawa et al. (2010) 
\enddata
 \tablenotetext{a}{
 For the CCSN models, the mass of the central compact remnant is not included. 
 The pre-SN wind velocity $V_\mrm{wind}$ and mass loss rate $dM/dt$ are fixed at 20 km~s$^{-1}$ and 10$^{-5}$ $M_\odot$~yr$^{-1}$ respectively, unless otherwise specified.
 }
\label{table:ejecta}
\end{deluxetable*}

\subsection{Post-shock Temperature Equilibration}
\label{section: temp}
We assume that, upon crossing the shock, thermal ions are heated to the same temperature per nucleon, i.e., all ion species obtain the same velocity distribution and $T_{i,2} = A_i \times T_{p,2}$.\footnote{We use subscripts `0', `1' and `2' to denote the far upstream, immediately upstream from the subshock, and the downstream regions respectively.} 
This assumption is consistent with particles making elastic scatterings off a massive background, as is generally assumed in DSA. Such behavior is also observed at the quasi-parallel Earth bow shock \citep*[][]{EMP90}.
For electrons, we assume that $T_{e,2}/T_{p,2} = m_e/m_p$ immediately behind the shock.

For simplicity, we also assume that $T_e/T_p = 1$ and $T_e/T_i = 1$ in the shock precursor as the particles are pre-heated by Alfv\'{e}n wave damping and adiabatic compression by the CR pressure. In fact, there is evidence for an upstream $T_e/T_p$ close to 1 from analysis and modeling of optical and X-ray emission from the shocks of, e.g., RCW 86 and Cygnus Loop 
\citep[e.g.,][]{Ghavamian2001,GLR2007,HelderVinketal2009,Helder2011}.

In each Lagrangian (i.e., mass) cell behind the shock, we follow the temperature evolution of electrons and each ion species by calculating their mutual equilibration (or heat exchange) rates. The equilibration is parameterized between the Coulomb collision rate (the slowest possible) and instant equilibration.
The Coulomb equilibration time-scale between two species of charged particles is computed using the following expression adapted from 
\citet{spitzer62}:
\begin{equation}
t_\mathrm{eq}(i,j) = \frac{3}{8\sqrt{2\pi}} \frac{m_i m_j}{n_j Z_i^2 Z_j^2 e^4 \log{\Lambda}} \left(\frac{kT_i}{m_i} + 
                                   \frac{kT_j}{m_j} \right)^{3/2}, 
\label{eqn:teq}
\end{equation}
where $i$ and $j$ stand for the equilibrating and target background particles respectively, $m$ is the particle mass, $n$ is the number density, $Z$ is the charge number, and $\log \Lambda = \log (b_\mrm{max}/b_\mrm{min})$ is the Coulomb logarithm for the given collision pair. Here $b_\mrm{max}$ is the Debye Length $\lambda_D = \sqrt{kT_e/(4\pi e^2 n_e)}$ and $b_\mrm{min} = \mrm{max}\{b_\mrm{min}^\mrm{cl}, b_\mrm{min}^\mrm{qm}\}$ \citep[e.g.,][]{Callen2006}. 
For electron-ion collisions, the classical limit $b_\mrm{min}^\mrm{cl} = Z_ie^2/(3kT_e)$ and the quantum limit $b_\mrm{min}^\mrm{qm} = h/(4\pi\sqrt{3kT_em_e})$, where $h$ 
is Planck's constant. For ion-ion collisions, we define the center-of-mass quantities $m_\mrm{cm} = m_im_j/(m_i+m_j)$ and $\overline{u^2} = 3(kT_i/m_i + kT_j/m_j)$ so that $b_\mrm{min}^\mrm{cl} = Z_iZ_je^2/(m_\mrm{cm} \overline{u^2})$ and 
$b_\mrm{min}^\mrm{qm} = 
h/(4\pi m_\mrm{cm}\sqrt{\overline{u^2}})$.
%
%

The temperature evolution for each species is then followed using an explicit finite difference scheme:
\begin{equation}
T_i(t+dt) = T_i(t) + \frac{T_j(t) - T_i(t)}{t_\mrm{eq}(i,j)}dt,
\end{equation}
where the time step $dt$ is much smaller than $t_\mrm{eq}(i,j)$.
For all models in this work, we consider equilibration through Coulomb collisions for which the time-scales are provided by Equation~(\ref{eqn:teq}), with the exception of one occasion for which we consider the extreme case of instantaneous equilibration established immediately behind the shock among all particle species.  

In a full calculation, one has to compute the equilibration in each Lagrangian 
grid cell and at every time step among a total of 140 charged particle species.
Each species can act as a target as well as an equilibrating particle at any given time.
This constitutes a $\{140 \times 140\}$ matrix for the equilibration network. To save computation time, we neglect a species in the equilibration network whose number density, relative to the most abundant species, is lower than some preset threshold value at a specific time-step 
and mass coordinate. For the results presented here, the threshold is $10^{-8}$ for an equilibrating particle and $10^{-3}$ for a target particle.
We have confirmed a posteriori that these threshold values 
result in X-ray spectra that do not differ noticeably from those obtained with no artificial limits.

Another simplification we make is that each ion species at a particular ionization state in a Lagrangian cell has a single temperature only; in reality, a mixture of temperatures in a ion species can be expected due to the dynamical nature of NEI combined with cross-species temperature equilibration in the post-shock plasma. This assumption allows us to easily follow $T_i$ at any space-time coordinate ($x$, $t$). 
A full treatment of this `multi-temperature' effect is reserved as a future task.

\subsubsection{A Test Calculation}

\begin{figure}
\centering
\includegraphics[width=7.5cm]{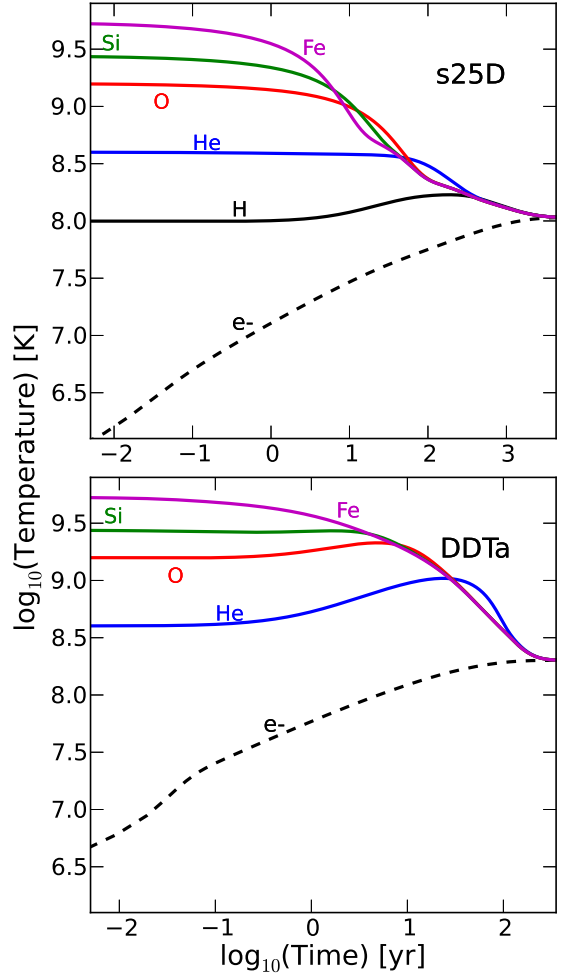}
\caption{
Test calculation of electron and ion temperature evolution in a Lagrangian cell with a fixed total ion density and ionization fractions. Full ionization is assumed here. The element abundance is taken from our s25D CCSN (upper panel) and DDTa Type Ia (lower panel) ejecta models by averaging over the ejecta volume weighted by the mass profile. In both cases, the ions begin with the same temperature per nucleon. Note that time is plotted in logarithm in order to show the detailed temporal behavior of the equilibrations more clearly. 
}
\label{fig:test_ion}
\end{figure}

We first consider a simple test problem in which electrons and ions are allowed to equilibrate in a Lagrangian grid cell with a constant total ion density $n_\mrm{ion}$, volume, elemental abundances, and ionization states.
The calculation starts at a time when the gas is shocked. We assume that the 12 elements we include (from H to Fe)
are fully ionized, and study two cases with different chemical abundances
taken from the s25D CCSN and DDTa Type Ia ejecta models described in the previous section. 
The abundance values are averaged over the ejecta volume weighted by the mass profiles provided by the corresponding SN explosion simulations. 
In Figure~\ref{fig:test_ion}, we show the results for a calculation with $n_\mrm{ion} = 100$~cm$^{-3}$. Under full ionization, the electron density is $n_e = 190$~cm$^{-3}$ for the CCSN case and $1930$~cm$^{-3}$ for the Type Ia case. The most abundant ion species of the s25D model are H and He with number fractions of 0.67 and 0.24 of all ions respectively, and Fe and He for the DDTa model with number fractions of 0.59 and 0.13 respectively. 
The initial proton and electron temperatures are set to $10^8$~K and $5.5\times10^4$~K respectively for both cases and the 
final adiabatic temperatures that all particle species reach in full equilibration are $1.1 \times 10^8$~K for s25D 
and $2.0 \times 10^8$~K for the Type Ia model.

Since $t_\mrm{eq}$ scales as $Z^2$, high-$Z$ ions, such as Fe$^{26+}$, equilibrate  
the fastest and rapidly approach the temperature of the most abundant element in the gas cell (note that $t_\mrm{eq} \propto n_j^{-1}$).
For the s25D model, all elements approach the H temperature (black solid curve in Figure~\ref{fig:test_ion}), while for the Type Ia model, all elements approach Fe (magenta curve).
 
It is noteworthy to contrast the behavior of H and He in the two cases.
In the CC s25D model, the temperature of the abundant He drops continuously with time, 
while the protons are slowly heated initially by interacting with the hotter He but then cool down slightly again as they reach equilibrium with 
He and now equilibrate with the heated electrons.
In the DDTa Type Ia model, because He and other lighter elements are less abundant than Fe, they first increase in temperature until they reach
quasi-equilibrium at $t \sim 10$\,yr. At this point, the evolution starts to be dominated by equilibration with the heating electrons and the ion temperatures drop toward the final $T_\mrm{adiab}$. 
The total time taken to reach full equilibration for the s25D case is longer than the DDTa case due to a lower metallicity and hence a smaller total number of particles in the grid cell.

In the s25D CCSN case, we observe that the heavy ion (i.e., C to Fe) temperatures first reach a quasi-equilibrium temperature a little higher than $T_p$ and lower than $T_\mrm{He}$ at $t \sim$ 100 yr. This indicates that there is an approximate balance between their equilibrations with H and with He which are the two most abundant elements in the gas. Then, $T_p$ and $T_\mrm{He}$, as well as the heavier ions, gradually converge toward $T_\mrm{adiab}$ by interacting with the electrons. 

While our simple test case shows expected equilibration behavior, the ion and electron temperature evolution in our nonlinear SNR models will be far more complex. In our simulations, the SNR is evolving under the influence of efficient DSA causing the overall plasma density to evolve 
in a non-self-similar fashion. The free electron density and the local ion charge states will be determined by NEI and recombination which depend on $T_e$, are coupled to the SNR evolution, and will vary with radius within the remnant. We know of no other SNR model combining as many physical effects in a self-consistent fashion.
%

\subsection{Spectral Line Broadening} 

Emission line profiles provide crucial information on the temperatures of the line-emitting ions from thermal broadening, and on the dynamics of the radiating gas from Doppler broadening due to gas expansion.
For the former, our self-consistent calculation of ion-ion and ion-electron equilibration with 
time- and space-dependent NEI, provides a reliable estimate of the heavy ion temperatures as a function of radius. 
The full-width-at-half-maximum (FWHM) dispersion of the Gaussian profile of an emission line centered at energy $E_\mrm{line}$ due to the thermal motions of the emitting ion $i$ is known to be:
\begin{equation}
\Delta E_\mrm{FWHM} \approx 2.355 \times \left(\frac{E_\mrm{line}}{c}\right) \sqrt{\frac{kT_i}{m_i}},
\end{equation}  
where $c$ is the speed of light in vacuum \citep[e.g.,][]{Rybicki79}. The time-evolution of the ion 
temperature gives the line profile a spatial (radial) dependence which, 
after projection along a line-of-sight (LOS),
can give rise to a non-trivial deviation from a single Gaussian profile.
This effect depends on the 
position of the LOS as it 
cuts through a different series of gas shells in the spherically symmetric SNR model. 

Our one-dimensional hydrodynamic code provides the velocity component of each gas element in the simulation parallel to the LOS to determine the Doppler shift of the observed line emission. Owing to the spherical symmetry, each line is split into two symmetric halves.  
The Doppler energy shift for the red- and blue-shifted lines is: 
\begin{equation}
\Delta E = \pm E_\mrm{line} \left( \frac{|v \cos \theta|}{c} \right),
\end{equation}
where $\theta$ is the angle between the velocity $v$ of the radiating gas element and the LOS.
Combined with thermal broadening, the Doppler shift can lead to a line being split into two, and sometimes closely packed lines blended together.
This is particularly important for a LOS close to the center of the SNR, where the expansion velocities of the shells of shocked gas can have larger components along the LOS compared to regions closer to the SNR rim.

\subsection{DSA at the Reverse Shock}
\label{section:dsa}
A detailed discussion of the \NL\ (NL) DSA calculation we use in the \crhydro\ code at the FS  is given in \citet*{LEN2012}. This generalized NL DSA calculation is again used here for the FS but, in this initial presentation of RS emission, we treat DSA at the RS in a simpler fashion. Since we do not include the non-thermal emission from DSA at the RS we do not need the full implementation for the FS used to model \SNRJ\ 
\citep[e.g.,][]{LEN2012}, Vela Jr \citep[][]{LSENP2013}, or Tycho's SNR \citep{Slane2014}.  
Instead, we use the simpler NL model discussed in \citet{EPSBG2007}.
The 2007 model, based on the \SA\ work of \citet*{BGV2005}, includes the basic NL effects of shock modification but does not have an explicit description of \MFA\ (MFA) or as detailed a description of the shock precursor as our 2012 implementation. Since DSA at the RS is still highly uncertain \citep*[see][]{EDB2005}, we leave a more detailed modeling of DSA at the RS to future work. 

An important approximation we make concerns the acceleration of heavy elements. As has been known for several decades 
\citep*[e.g.,][]{Eichler79,EJE1981,EMP90}, high mass to charge ratio ions will be preferentially accelerated compared to protons in \NL\ DSA. While this effect can be treated exactly with \mc\ methods \citep[e.g.,][]{JE91} and plasma simulations \citep[e.g.,][]{SKT99}, it must be parameterized in \SA\ calculations.

\citet*{CBA2011} have included the enhancement of 
heavy ions in recent versions of their NL DSA model 
but we have not yet implemented this in the \crhydro\ code. 
Instead, we have simply included the effect of helium on the \gamray\ production with a scaling factor. While this approximation is adequate for the FS interacting with an ISM composed primarily of hydrogen, it would not be a good approximation for \gamray\ production at the RS where the ejecta material can be dominated by elements much heavier than hydrogen.

Instead of implementing a full $A/Z$ enhancement calculation, when considering particle acceleration at the RS we simply assume that the ejecta mass is composed of hydrogen and use the DSA calculation of \citet{EPSBG2007} to obtain the NL effects on the post-shock temperature and density which influence the thermal X-ray emission.

Another important, but poorly understood property of the RS concerns the magnetic field strength the RS encounters, $B_\mathrm{RS}$. Any seed magnetic field from the pre-SN star will be diluted by expansion to values considerably less than typical ISM values. These values 
(e.g., $B_\mathrm{RS} \lesssim 10^{-9}$\,G) may not even be large enough to allow a thin collisionless RS to form suggesting that significant MFA occurs at the RS \citep[e.g.,][]{EDB2005}. While MFA is calculated assuming a resonant wave$-$particle interaction at the FS in \crhydro, for the RS we use the simpler ad hoc MFA as described fully in \citet{EC2005}. Typically we use an ad hoc MFA parameter that yields an unshocked field 
immediately upstream from the RS of $B_0 = 0.1$\,\muG. 

While we include effects of efficient DSA at the RS on the thermal X-ray emission from the ejecta material in some of the models we discuss below, we leave a detailed calculation of RS non-thermal emission from radio to gamma rays to a 
future study. As indicated above, this will require including heavy ion acceleration and enhancement in the \crhydro\ code.

\subsection{Parameter Setup}

We consider an SNR evolving in a 
spherically symmetric ambient medium with a set of 
environmental and model parameters typical of young Galactic SNRs. 
Parameters that remain fixed for all of our models are: 
\begin{enumerate}[(i)]
\item an age of 500\,yr at a distance of 3\,kpc from Earth; 
\item cosmic abundances  in the ambient medium;\footnote{We have ignored the possible mix of heavy elements in the pre-SN wind for some CCSN models here for simplicity.} 
\item  a constant ambient magnetic field strength of $B_0=3$\,\muG~for Type Ia SNe, and a constant magnetization parameter \citep[e.g.,][]{CL94} of $\sigma_\mrm{wind} = 0.03$ for a pre-CCSN wind; 
\item an electron-to-proton number ratio at \rel\ energies, $\Kep = 0.01$;
\item a pre-shock ambient temperature $T_0 = 10^4$\,K; and, 
\item a free-escape-boundary at 10\% of the shock radius ahead of the FS at which the highest momentum particles escape from the shock. 
\end{enumerate}
Quantities that are varied in some models include: 
\begin{enumerate}[(i)]
\item the pre-shock gas density in the ambient medium, namely the ISM density $\nISM$ for Type Ia models and the pre-SN wind parameters for the CCSN models;
\item  a parameter, $\fDamp$, characterizing the rate of \Alf\ wave damping in the shock precursor that 
pre-heats the in-flowing gas; 
\item  the acceleration efficiency of NL-DSA at the FS, $\EffDSAfs$, including MFA, and at the RS, $\EffDSArs$, with a fixed ad hoc MFA;  
\item the rate of temperature equilibration among charged particles behind the shocks; in Section 3.7, we carry out one calculation with fast equilibration against pure Coulomb interactions (Equation~(\ref{eqn:teq})) to study its role in shaping the X-ray spectrum.
\end{enumerate}
For detailed definitions of these parameters, and a description of the underlying physics formulation in the simulation, see e.g., \citet[][]{LEN2012}. 

We include the 12 most abundant elements in our NEI and thermal X-ray calculations: H, He, C, N, O, Ne, Mg, Si, S, Ar, Ca, and Fe. 
Once the simulation is finished, the temperatures and ionization fractions for all charge states of these elements is known for each thin spherical shell of material between the RS and the FS. This information allows us to calculate the volume integrated and LOS thermal X-ray line spectra as a post-processing step. 

\subsection{Numerical Setup and Performance}

Before moving on to the results, we first provide the basic specifications of the \textit{CR-hydro-NEI} code and describe its overall performance.  
The total number of Lagrangian cells used in the simulation domain is fixed at 3000. The location of the outer boundary of the domain is adjusted according to the final position of the FS at 500~yr for each model. The X-ray spectra are calculated with 10000 uniformly-spaced energy bins spanning from 95~eV to 12~keV with a bin size of 1.2~eV (c.f.~the design energy resolution of the \textit{Soft X-ray Spectrometer} (SXS) onboard \textit{Astro-H} is 7~eV or better from 0.3 to 12~keV). To calculate the synchrotron emission, the shocked gas volume between the CD and FS is split into a few tens of concentric shells in which the evolution of the non-thermal electron distributions inside is followed, with synchrotron, Coulomb, adiabatic and various other energy losses taken into account \citep[for details of this method; see, e.g.,][]{LKE2008}. The hydro time step is limited by the Courant condition on the sound or shock crossing time (whichever is the shorter) for the smallest grid, or the shortest timescale associated with e.g. radiative cooling which is not important for the models presented here.

Under these specifications, the serial code can run efficiently on a reasonably fast CPU onboard a modern laptop computer. For example, it takes typically shorter than 5~hr to finish a 500~yr simulation on a 2.4 (3.4 boosted)~GHz Intel Core i7 CPU, and can be faster depending on the problem. The memory requirement is very modest considering it is a one-dimensional code and is not a limiting factor.

\section{Results and Discussion}
\label{result}
Our primary focus in this initial presentation of our RS/FS model is to demonstrate the capability of our generalized \textit{CR-hydro-NEI} code to perform self-consistent calculations of X-ray spectra for different circumstellar environments and progenitor/ejecta models using data adapted from SN simulations with explosive nucleosynthesis. 
Besides current observations, we hope to demonstrate the usefulness of \crhydro\ for modeling 
future high-resolution X-ray spectroscopic observations 
of various types of young SNRs by 
the next-generation space telescopes, such as \textit{Astro-H}.
%

In Section 3.1$-$3.3, we discuss and compare general results for different SN and progenitor models based on a set of preset values for influential parameters defined in Section 2.5: $\fDamp = 0.1$, $\EffDSAfs = 0.7$, $\EffDSArs < 0.01$, $n_\mrm{ISM} = 0.2$~cm$^{-3}$ for Type Ia models and $dM/dt = 10^{-5}$~$M_\odot$~yr$^{-1}$, $V_\mrm{wind} = 20$~km~s$^{-1}$ for the pre-SN wind in the CCSN models. Temperatures of the charged particles are equilibrated by Coulomb interactions. In Section 3.4$-$3.7, we vary these parameters one-by-one and study in detail their effects on the X-ray emission properties. Finally, in Section 3.8, we look at the time evolution of the thermal X-ray spectra.   

\subsection{Ionization in Shocked Plasma}
 
\begin{figure*}
\centering
\includegraphics[width=15cm]{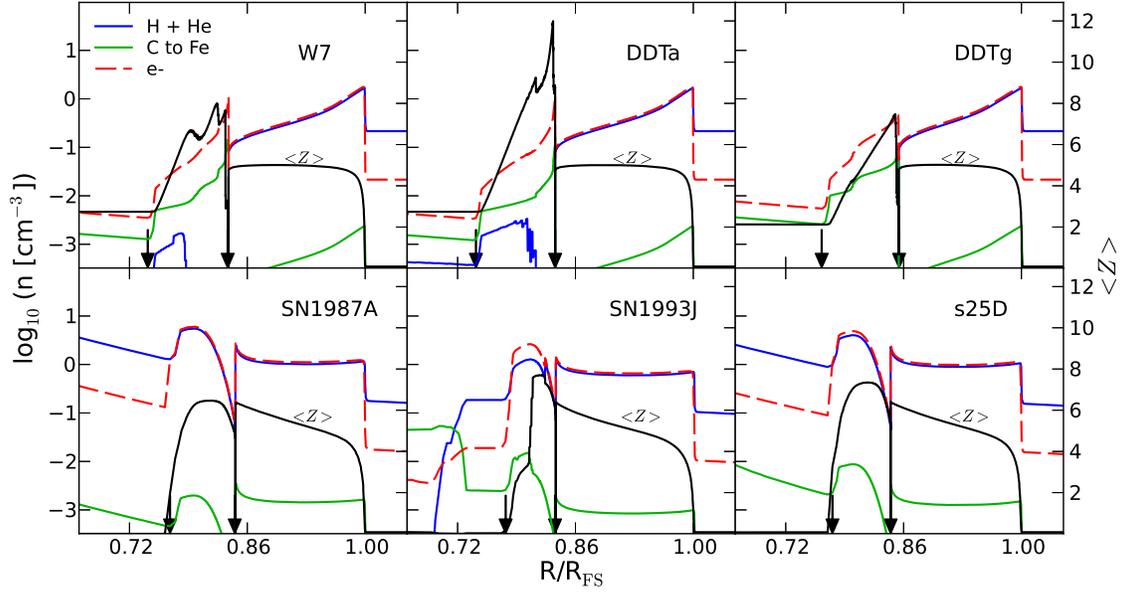}
\caption{
Number density profiles of $e^-$ (dashed line) and ions (colored solid lines) from the RS to the FS at 500~yr for the 
Type Ia (top three panels) and CCSN ejecta models (bottom three panels). 
The black solid lines scaled by the right-axes show the number-averaged charge number $\Zave$ of the plasma (excluding H and He ions which do not contribute to the X-ray emission), which visualizes the overall level of ionization of the gas as a function of radius.
Here, and in subsequent figures, the arrows indicate positions of the RS and CD at 500~yr. 
For all six cases, $\fDamp=0.1$, $\EffDSAfs=0.7$, and $\EffDSArs < 0.01$. For the Type Ia models in the top three panels, $\nISM =0.2$\,\pcc, while for the wind models in the bottom panels, $dM/dt=10^{-5}$\,\SunMyr\ and $\Vwind=20$\,\kmps.
(A color version of this figure is available in the online journal.)
}
\label{den_tot_profile}
\end{figure*}

We first show in Figure~\ref{den_tot_profile} the density profile of free electrons ($n_e$) and ions ($n_\mrm{ion}$, separated into two elemental groups of H+He and C to Fe) in the RS-CD and CD-FS post-shock regions after 500~yr of evolution, where CD stands for contact discontinuity, the radius separating the shocked ejecta from the shocked ISM/CSM. 
To roughly indicate the ionization state of the X-ray-emitting plasma, we also show 
the number-averaged charge number $\Zave$ as a function of radius. 
We define $\Zave \equiv \sum_{i,j}{n_{ij}Z_{ij}}/\sum_{i,j}{n_{ij}}$, where $n_{ij}$ and $Z_{ij}$ are the number density and charge of an ion species of element $i$ and ionization state $j$. In calculating $\Zave$, H and He are excluded since they do not contribute to the X-ray emission.

The more massive ejecta of the CCSN models (especially SN1987A and s25D; bottom panels in Figure~\ref{den_tot_profile}), as well as the overall slower expansion rates of their remnants due to the encounter of the ejecta with a dense red supergiant (RSG) wind,
lead to a much higher total density in the RS-CD region than in the Type Ia models. This contrast in density can lead to significant differences in the particle heat exchange and ionization rates, and consequently in the X-ray spectra and line profiles.

For the Type Ia models, $n_e$ dominates over $n_\mrm{ion}$ in most of the RS-CD region since the plasma there is generally heavy ion rich. In this case, it can be shown that $\Zave$ is roughly equal to the electron-to-ion number ratio $n_e/n_\mrm{ion}$. 
We can also see that, due to the relatively low density and high metallicity of the Type Ia ejecta, $n_e$ rises much more slowly against $n_\mrm{ion}$ behind the RS than in the CCSN ejecta. 
Some structures can also be found in the $\Zave$ profiles of the W7, DDTa and SN1993J models in the RS-CD region. These are caused by the transition of chemical compositions experienced by the RS as it propagates through the ejecta layers. 

In all of our CCSN models, the RS positions at 500~yr are exterior to the inner ejecta layers where most of the heavy elements are located. As a result, the shocked ejecta material is mostly composed of H and/or He, and a full ionization is realized almost immediately behind the RS. This is reflected by the quick saturation of the $n_e/n_\mrm{ion}$ ratio in the RS-CD region. A similar fast saturation of $n_e/n_\mrm{ion}$ is also observed in the shocked ISM/CSM for all six models since the FSs are propagating in an ambient medium with the ISM composition.

The ionization states of the heavy elements, indicated by $\Zave$, are quite different between the Type Ia and CCSN models in the CD-FS region. The Type Ia cases all show a quick rise of $\Zave$ behind the FS followed by a flat spatial profile, whereas the CCSN cases show a more gradual increase behind the FS, followed by a mild gradient rising toward a maximum at the CD. This difference originates from the fact that the FSs are propagating into a pre-SN wind in the CCSN models against a uniform ISM in the Type Ia models. The mild gradients of $\Zave$ in the CCSN cases stem from the $\rho \propto r^{-2}$ density profile in the wind, while the more gradual rise of $\Zave$ behind the shock is because of the lower density of the wind that the FS are propagating into at 500~yr  than the uniform ISM in the Type Ia models.  

\begin{figure*}
\centering
\includegraphics[width=14cm]{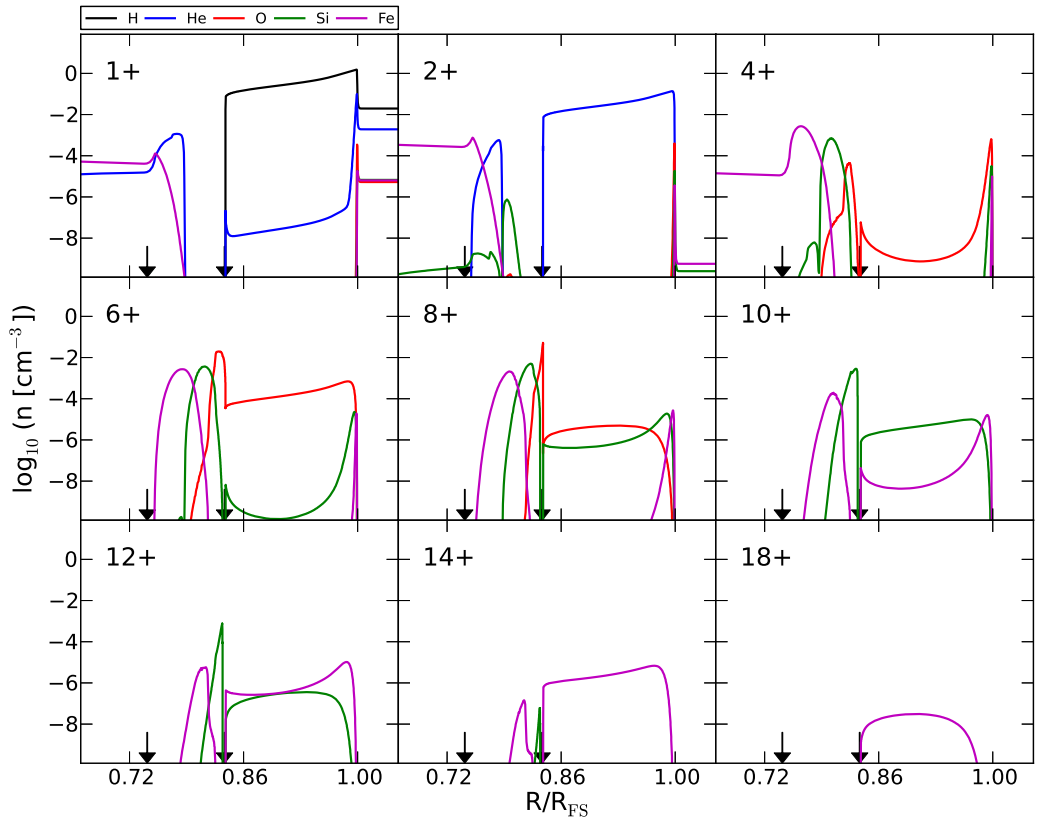}
\caption{
Number density profile of individual ions at 500~yr for the W7 ejecta model shown in Figure~\ref{den_tot_profile}. Each panel corresponds to the specified ion charge state. (A color version of this figure is available in the online journal.)
}
\label{ion_profile_ddta}
\end{figure*} 

To better see the composition and ionization state of the shocked plasma, Figure~\ref{ion_profile_ddta} shows the density profiles for a set of selected ion species for the W7 ejecta model shown in Figure~\ref{den_tot_profile}. For each species, the profiles are displayed from the +1 to +18 ionization states. We can immediately see that the CD-FS side has a much simpler set of profiles than what we observe in the RS-CD region, since the FS is running into a uniform ambient medium with a fixed, spatially independent cosmic abundance, whilst the RS has propagated into a more complicated structure in the ejecta material as shown in Figure~\ref{fig:ccsne} and \ref{fig:type1a}. 
For the RS side, the resulting profile of each particular ion species depends on the changing composition in the shocked ejecta layers in addition to the ionization time. Species of higher ionization states generally concentrate at regions closer to the CD since those particles have experienced a longer ionization time than those closer to the shocks. An exception to this results if a particular ion species has a low overall molar fraction in the outer layers of the ejecta that can hence ``delay'' their appearance in the final profiles until later times, e.g., He$^{2+}$. This interesting interplay between the ejecta structure and NEI can also be seen in the profiles of O ions for example, and it can have some consequences on the X-ray morphology of young SNRs in different energy bands.

\begin{figure*}
\centering
\includegraphics[width=14cm]{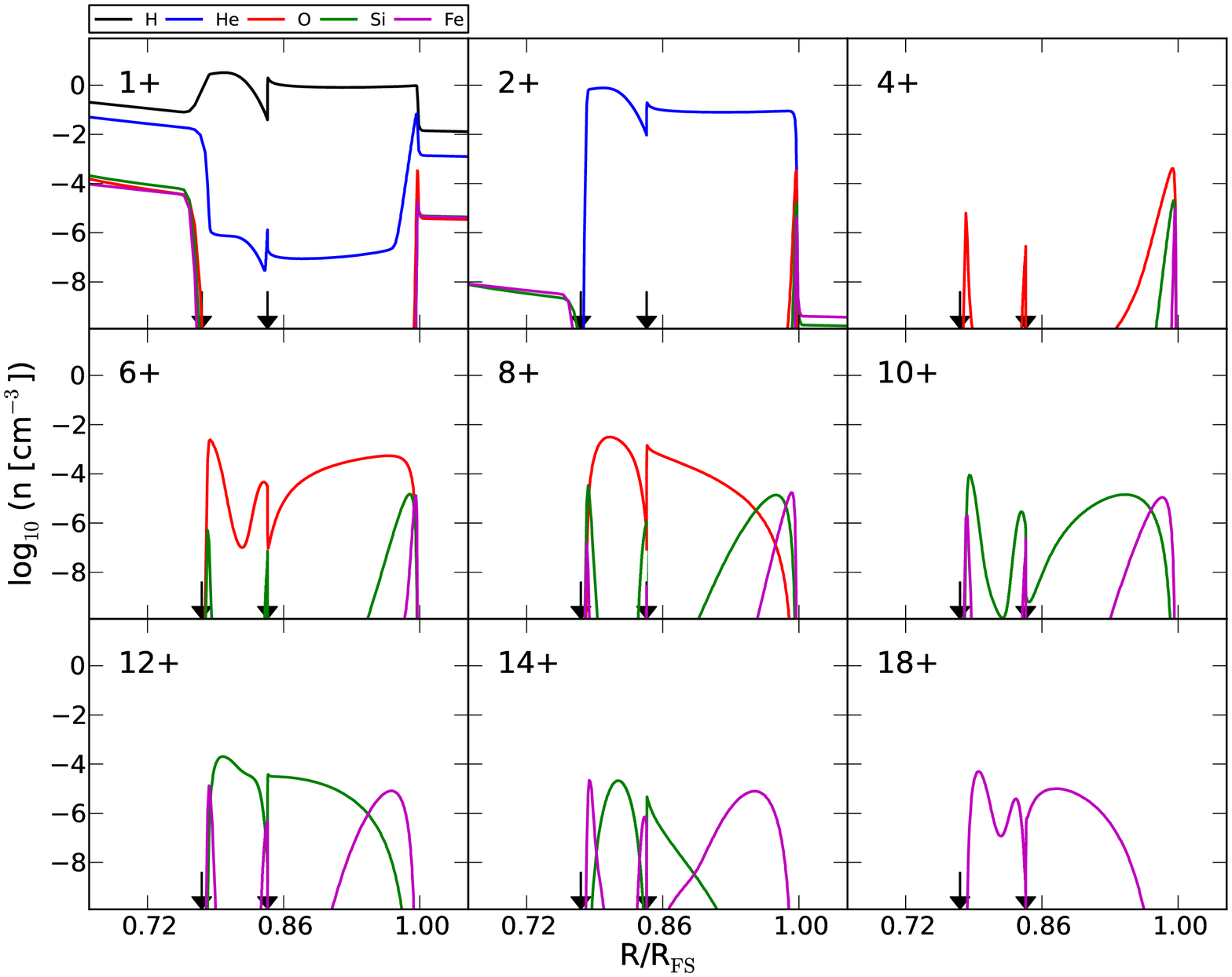}
\caption{
Same as Figure~\ref{ion_profile_ddta} but with the s25D CCSN ejecta model 
shown in Figure~\ref{den_tot_profile}.
(A color version of this figure is available in the online journal.)
}
\label{ion_profile_s25d}
\end{figure*}

Figure~\ref{ion_profile_s25d} shows similar plots for the s25D CCSN ejecta model shown in 
Figure~\ref{den_tot_profile}.  
The RS and CD radii relative to the FS are changed due to the very different initial ejecta dynamics. The effect of the FS running into a density gradient of a pre-SN wind can be observed in the ion density profile, i.e., ions of higher ionization states are generally more concentrated in the inner part closer to the CD where the FS encountered a denser wind. Behind the CD, since the outer envelopes of this CCSN ejecta have a much more similar composition to the upstream CSM than the Type Ia case, the contrast between the RS-CD and CD-FS regions is a lot different from what we see in Figure~\ref{ion_profile_ddta}. The most obvious factor in play here is the fast adiabatic expansion of the ejecta material (i.e., the density of the gas is decreasing quickly as the SNR ages) which does not affect as much the FS-shocked CSM. 
This adiabatic effect dilutes the shocked ejecta gas, the ionization age $n_et$ remains small, and the ionization can saturate at intermediate states with the mass fractions of highly-ionized species remaining low, despite a relatively long elapsed time since being shocked.

Plasma shocked earlier on in the SNR evolution (i.e., plasma closest to the CD in the ejecta side) suffers the most from this effect since the SNR expansion was the fastest back then when it was shocked. 
We can see that the density profiles of intermediate ions such as Si$^{4+}$ to Si$^{6+}$ show an enhancement near the CD while the densities of higher ionization states remain low. This is different from the behavior of the plasma further inward toward the SNR center and can be interpreted as the result of this `quenched ionization'-like effect. It is also observed in the Type Ia case in Figure~\ref{ion_profile_ddta}  (see, e.g., the peaky features of the intermediate O ions), albeit less prominently mainly due to the relatively low ejecta density and hence slower ionization throughout the shocked ejecta region. Similar behaviors have been discussed in some previous works on NEI plasma in young SNRs \citep[e.g.,][]{badenes03, Badenes2005}. It hence serves to verify that our calculations are correctly following the dynamical coupling between the NEI states of heavy elements and the hydrodynamics.

\begin{figure*}
\centering
\includegraphics[width=16cm]{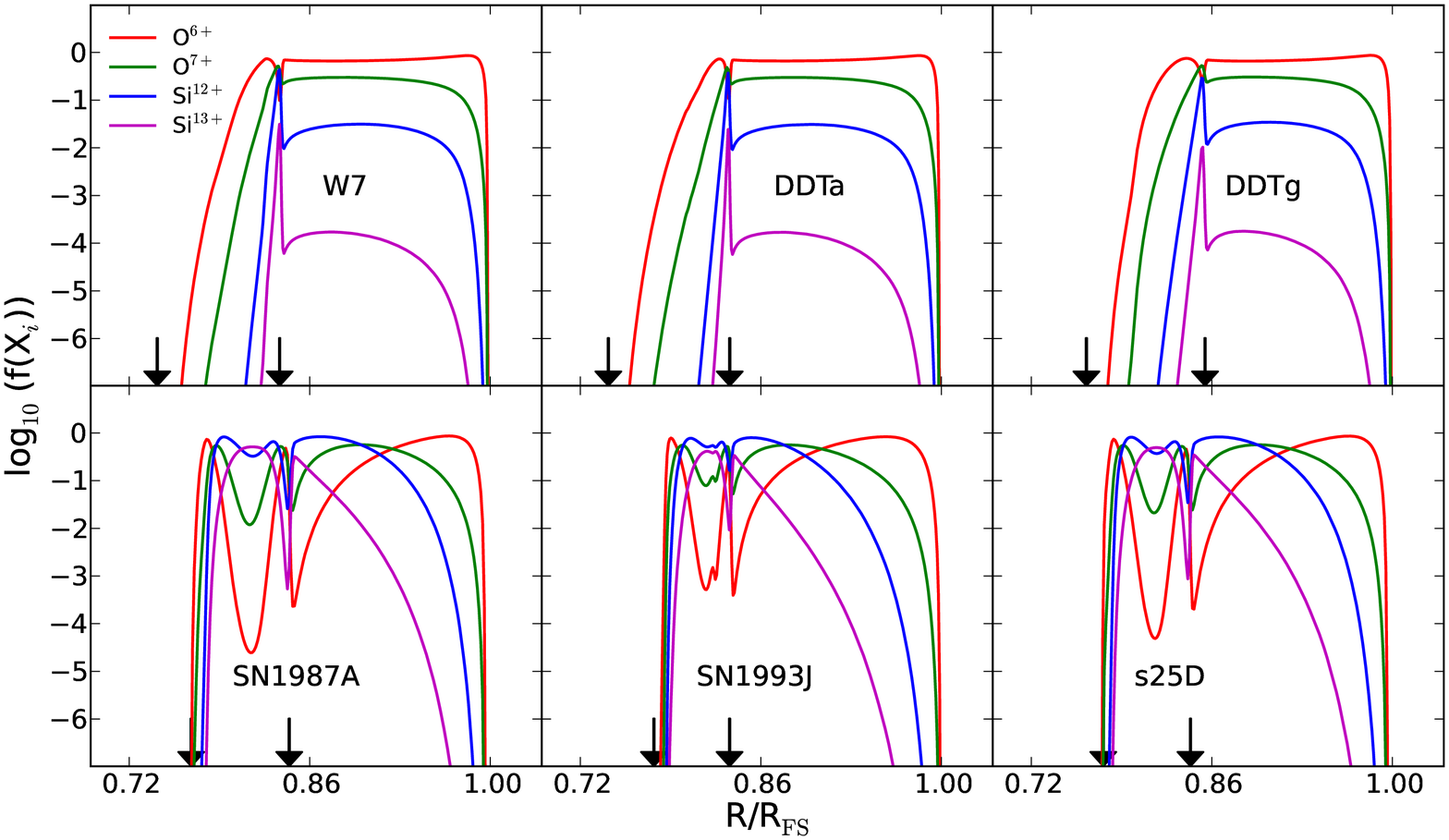}
\caption{
Number fraction $f(X_i)$ of H-like and He-like O and Si ions at 500~yr for each of the ejecta models shown in Figure~\ref{den_tot_profile}.
(A color version of this figure is available in the online journal.)
}
\label{ion_fraction_O_Si}
\end{figure*}

It is also instructive to examine the spatial distribution of the relative ionization fraction of important X-ray-emitting ions at the end of the simulation, such as those often used for plasma density and temperature diagnostics in X-ray spectroscopy. As an example, number fractions $f(X_i)$ of H-like and He-like ions O$^{6+}$, O$^{7+}$, Si$^{12+}$ and Si$^{13+}$ as a function of radius 
are shown in Figure~\ref{ion_fraction_O_Si}. Clearly, the differences among models from the same progenitor class are far less than the differences between classes.
In the shocked ejecta, higher ionization states like those shown in the plot arise much closer to the RS in the CCSN models. This is mainly attributed to the much higher shocked ejecta density in these models than the Type Ia cases at any given time in the simulation. The spatial structures seen in $f(X_i)$ of the shocked ejecta are again interpretable as being due to the ``quenched ionization'' effect from adiabatic expansion mentioned above; H-like Si is found mostly in the middle between the RS and CD, while He-like Si shows a concave structure peaking at both ends. Both H-like and He-like O show this concave behavior since oxygen is almost fully ionized in the middle part for our CCSN models. 
A small structure also appears behind the CD in the ejecta of the SN1993J model, which is due to the transition from the thin H-rich envelope to the He-rich region.      

In the CD-FS region, interesting differences between the Type Ia and CCSN cases due to the presence/absence of a density gradient from a pre-SN wind are also observed.
For the Type Ia models evolving in a constant density ISM, the H-like to He-like ion number ratios rise gently (but always $< 1$, for the ambient density we consider here)
from the FS to the CD.  These ratios show a much more drastic change in radius for the CCSN models evolving in a wind. Here, the FS propagates in a dense wind initially and moves into lower density material later on. This results in much higher H-like to He-like ratios, which, for the parameters we use here, can be $>1$ for lighter elements like O, closer to the CD.

These explicitly calculated ion density profiles, after combined with temperature and bulk motion information that we discuss in the next sections, are used directly to synthesize thermal X-ray spectra from the RS and FS regions.  

\subsection{Temperature Profiles}
\begin{figure*}
\centering
\includegraphics[width=16cm]{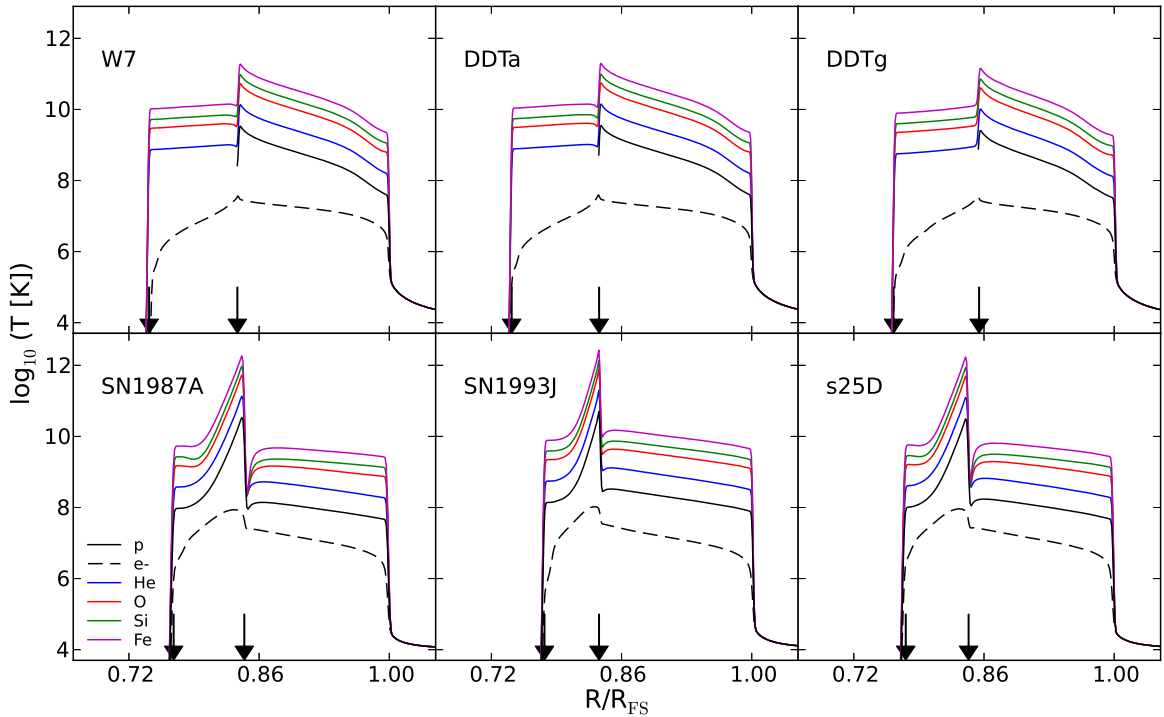}
\caption{Temperature profile of electrons, protons and heavy ions at 500~yr for each ejecta 
model shown in Figure~\ref{den_tot_profile}.
Temperatures of the heavy ions are averaged values weighted by the number density of each charge state of the same element (see text).  
Since only a slow equilibration by Coulomb interactions is assumed in these models, the ion temperatures are not significantly influenced by the equilibration after being shocked. 
(A color version of this figure is available in the online journal.)
}
\label{temp_profile}
\end{figure*}

The temperature of each particle/ion species ($T_e$, $T_p$, $T_i$) at each spatial grid and time step is kept track of by the calculation described in Section~\ref{section: temp}. Figure~\ref{temp_profile} shows the final temperature profiles of the downstream plasma from the RS to the FS for our Type Ia and CCSN models shown in Figure~\ref{den_tot_profile}. 
The ion temperatures shown here (from He to Fe) at each spatial grid are average values  $<$$T_i$$>$ weighted by the number fraction $X_{ij}$ of all charge states $j$ for each element $i$ of which the temperature is $T_{ij}$. In other words, we define $<$$T_i$$>$ $\equiv \sum^{}_{j}X_{ij} T_{ij}$ where $\sum^{}_{j} X_{ij} \equiv 1$ for all $i$. 


Overall, the CCSN models show a flatter temperature profile for the ions in the CD-FS region than their Type Ia counterparts. The major reason is that the shock velocity, 
which is the most important determining factor for the shocked temperatures, has a much flatter time evolution for the CCSN cases where the FS is propagating in a decreasing density gradient inside the pre-SN wind and hence suffers less deceleration.
The initial shocked temperature in the early phase (close to the CD) is higher for the Type Ia models since the FS propagates in a dense wind in the beginning for the CCSN cases and hence has a lower initial velocity. 

The difference in the initial energetics and mass profile of the ejecta between the Type Ia and CCSN models also contributes to the difference in the shock dynamics and hence the temperatures in the CD-FS region. In front of the FS, a ramp due to pre-heating by damping Alfv\'{e}n waves (self-generated by the accelerated protons through the streaming instability) can also be observed. We study the effect of strong damping versus weak damping in the CR precursor on the emission in a following section. 

For the RS-CD region, the profiles and absolute temperatures for the two SN classes also have distinct behaviors. 
In the CCSN models, the ejecta initially encounter a dense wind and a strong RS is created with a high velocity in the ejecta rest frame. This results in high shocked temperatures immediately behind the CD.
However, the RS velocity slows rapidly as it propagates through the dense ejecta producing a steep temperature gradient in the RS-CD region.
The Type Ia models exhibit much flatter temperature profiles since the RS evolves in lower density ejecta and decelerates less rapidly.
The very different chemical composition in the shocked ejecta material also plays an important role in determining the absolute plasma temperature. 

For these models, we found that the temperatures of heavy ions do not exhibit any strong heating/cooling from interaction with other elements in either the CD-FS or RS-CD regions.
%
This is because Coulomb equilibration is assumed for heat exchange which results in an equilibration time long compared to the dynamical time-scale for these models. In Section 3.7, we will present a model with an instantaneous temperature equilibration to show the impact on the resulted X-ray line emission.
 
We note that ionization cooling of electrons is not considered in our calculations. If the electron temperature immediately behind the shock is orders of magnitudes higher than the mass-ratio value (i.e., $T_e \gg (m_e/m_p)\times T_p$), ionization cooling can impose important effects on the electron temperature profile close to the shock, especially on the ejecta side. Nevertheless, in our models where the mass-ratio value is adopted, ionization cooling is expected to be much less important, and the electron temperature evolution behind the shock is dominated by collisional heating \citep[e.g., see Figure~7 of][]{Yamaguchi2014}. 

\subsection{X-ray Spectra}  

\subsubsection{Volume-integrated Spectra}
\begin{figure*}
\centering
\includegraphics[width=16cm]{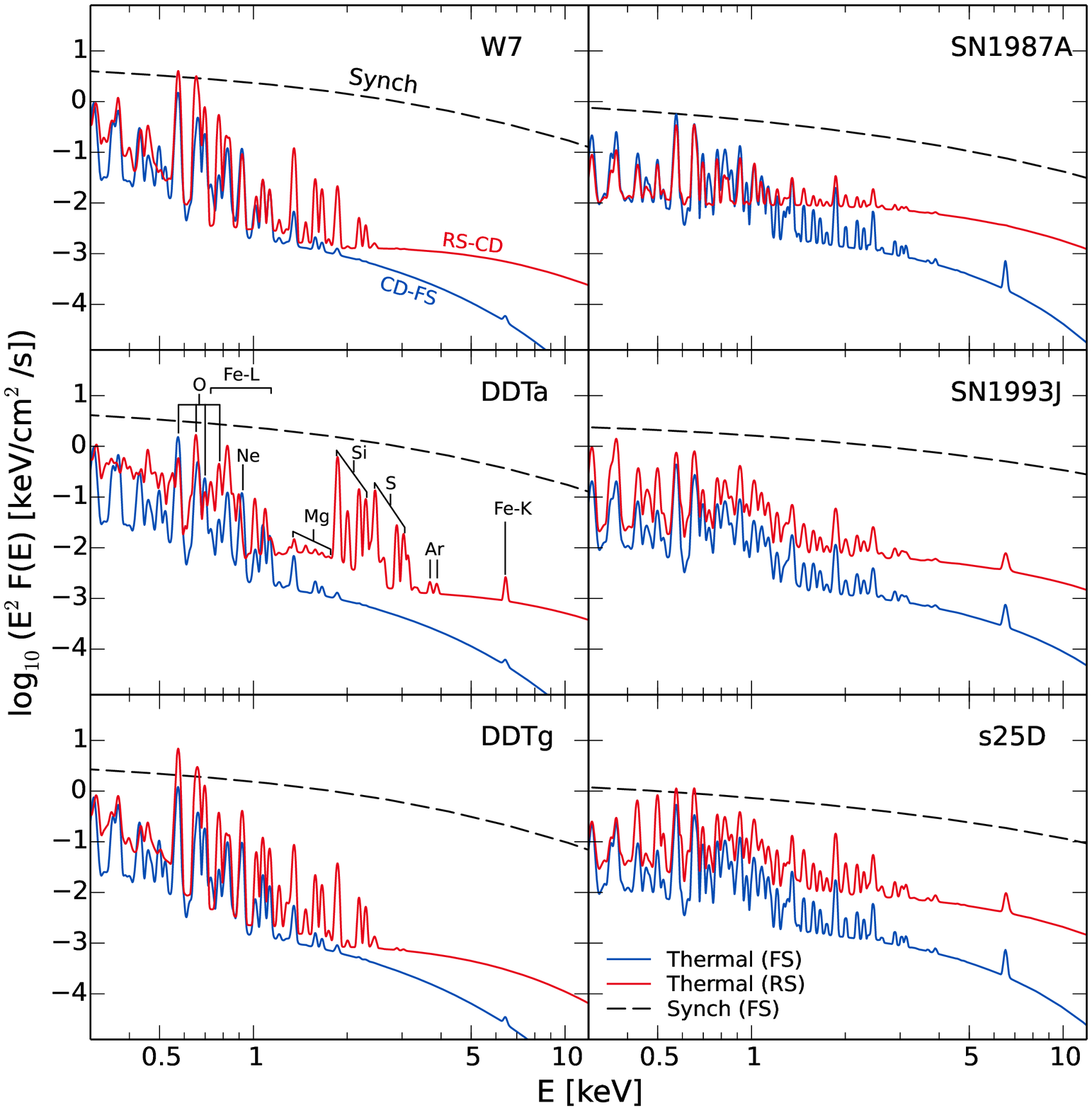}
\caption{
Volume-integrated X-ray spectra (CD-FS and RS-CD regions) from 0.3 to 12~keV at 500~yr for the Type Ia and CCSN ejecta models. The thermal spectra are convolved with a Gaussian with FWHM of 7~eV, the baseline energy resolution of the SXS instrument onboard \textit{Astro-H}. The intrinsic line profiles are calculated including thermal broadening using the ion temperatures as well as Doppler shifts using the bulk velocity of the radiating gas from the simulation. Some of the stronger lines are labeled in the panel for the DDTa model. Lines with centroid energies below the O lines (not labeled) are mostly emitted by light elements like C and N, or heavier elements like S and Ca in low ionization states. The non-thermal spectra (from the CD-FS region only) are also plotted for reference.}
\label{spec_integ}
\end{figure*}

After obtaining the number densities, ionization fractions and temperature information for the 140 ion species in the shocked gas parcels, the local thermal X-ray continua and line spectra can be readily calculated as a post-process. 
In Figure~\ref{spec_integ}, we show volume-integrated thermal spectra in a broad energy range of $0.3 - 12$~keV from the shocked ejecta (red curves) and shocked ISM/CSM (blue curves). 
Doppler broadening from thermal motion, as well as from the bulk plasma flow, is taken into account here and in all subsequent results. 

First of all, the CD-FS spectra from different models of the same progenitor class are similar since we have assumed cosmic abundance in the ambient medium for all models. Any differences in the spectra from the shocked ISM/CSM originate from the SN energetics 
and/or the density structure.
The density structure of the pre-SN wind is responsible for the stronger high-energy lines from the CD-FS spectra in the CCSN models than in the Type Ia cases (blue curves in Figure~\ref{spec_integ}).
These lines originate from high ionization states of heavier elements like Si, S, Ar, Ca and Fe. These high ionization states are mainly generated in the early phase of the evolution when the FS is pushing through a high density RSG wind.
The thermal X-ray continua from the CD-FS region are also found to extend to higher energies in the CCSN models since the number-averaged $T_e$ in the post-shock gas are roughly $30$\% $-$ $50$\% higher than those in the Type Ia models, which also originate from the different hydrodynamic evolution of the FS and shocked gas in a pre-SN wind against a uniform ISM.

On the other hand, the larger variety of ejecta masses for the 
CCSNe models give rise to a dispersion in the shock dynamics, as indicated by the final FS radii and velocities of the s25D, SN1987A and SN1993J models 
(i.e., 2.8, 2.6, 3.4~pc and 4000, 3700, and 4900\,\kmps, respectively).
This variety produces different shocked ion temperatures and different levels of thermal line broadening in the CD-FS spectra.  

For the RS-CD spectra, the composition structure of the shocked ejecta comes into play.
For example, strong H-like O lines are produced in the DDTg and W7 ejecta, while the S, Ar, Ca and Fe K lines are obviously much weaker than in the DDTa ejecta. 
It is clear from Figure~\ref{fig:ccsne} that for the CCSN models, except the low-mass SN1993J case, the RSs are unable to penetrate through the massive outer H-rich envelope of the dense ejecta in 500\,yrs. Because of this, the composition of the shocked ejecta material resembles the shocked ambient gas with only small fractional differences. Thus, depending on the CCSN model and the ambient density, the X-ray line spectra from the forward and RSs can be similar for 100s of years.

In the SN1993J model, the RS did manage to propagate back to the He and C+O rich layers of the Type IIb ejecta. However the RS is still well separated from the inner heavy element rich core.
The total masses of shocked ejecta for the Type Ia models are roughly in the range of $0.7-0.8$~$M_\odot$, indicating that the dynamics of the RS in these models are similar. This range is more diverse for the CCSN models, roughly $1.0 - 1.4$~$M_\odot$, again owing to the larger variety of their ejecta masses.    

In young SNRs, thermal emission is likely to be accompanied by strong X-ray synchrotron emission from \rel\ electrons accelerated at the FS. 
We show the non-thermal synchrotron spectra from the CD-FS region (dashed curves) in Figure~\ref{spec_integ}.
The synchrotron component in these models is strong relative to the thermal emission for three main reasons.
First, we use a high acceleration efficiency at the FS, $\EffDSAfs=0.7$ for these models, 
to emphasize thermal X-ray emission under efficient DSA which may be higher than is typical.
For example, the most recent model of the young Type Ia SNR Tycho has inferred $\EffDSAfs=0.26$ \citep{Slane2014}. 

Second, we do not consider the possible fast damping of the amplified $B$-field behind the shock that has been suggested to be happening at some young X-ray synchrotron emitting SNRs \citep[e.g.,][]{Pohl2005, CassamEtal2007}. An effectively damped $B$-field behind the FS would certainly suppress the total X-ray synchrotron flux. It may also be possible that the damping process deposits heat to the shocked gas which, if it is fast enough and the amplified field strength is large, can affect the post-shock temperature and hence the ionization states of the plasma. 

Lastly, we have fixed $\Kep$ at 0.01 and lower values would again reduce the X-ray synchrotron emission. 
Our understanding on the process of electron injection in DSA at strong collisionless shocks is still incomplete. Recent kinetic/hybrid simulations have started to shed light on the complex wave-particle interactions that are responsible for facilitating electron injection \citep[e.g.,][]{RiqSpit2011}; these interactions may pre-accelerate the thermal electrons to super-thermal energies to reproduce an injection fraction that is compatible with $\Kep \sim 0.001 - 0.01$ suggested by CR measurements \citep{Kang2014}. Nevertheless, before we can determine $\Kep$ at strong SNR shocks from first principles with more confidence, a full utilization of multi-wavelength data (including X-rays) from an SNR is necessary to constrain $\EffDSA$ and $\Kep$ simultaneously \citep[e.g.,][]{LSENP2013}. Since we do not consider multi-wavelength modeling here, the relative normalizations between the synchrotron and thermal spectra are beyond the scope of this work and will not be discussed further.

\subsubsection{Spatially resolved Spectra along Line-of-sights}
Next we show LOS spectra which are required if a remnant can be spatially resolved by an X-ray spectrometer.
For reference, using typical parameters for the ambient medium, the models shown in Figure~\ref{den_tot_profile}
yield an angular diameter for a 500~yr old SNR of $\Delta\theta \sim 20'$ ($d_\mrm{SNR}/1$~kpc)$^{-1}$ for the CCSN+RSG wind cases, and $\sim 30'$ ($d_\mrm{SNR}/1$~kpc)$^{-1}$ for the 
Type Ia cases. Thus, an instrument with an angular resolution of a few arcmin could resolve nearby remnants unless they are substantially smaller due to the interaction with dense material such as molecular clouds.

\begin{figure*}
\centering
\includegraphics[width=16cm]{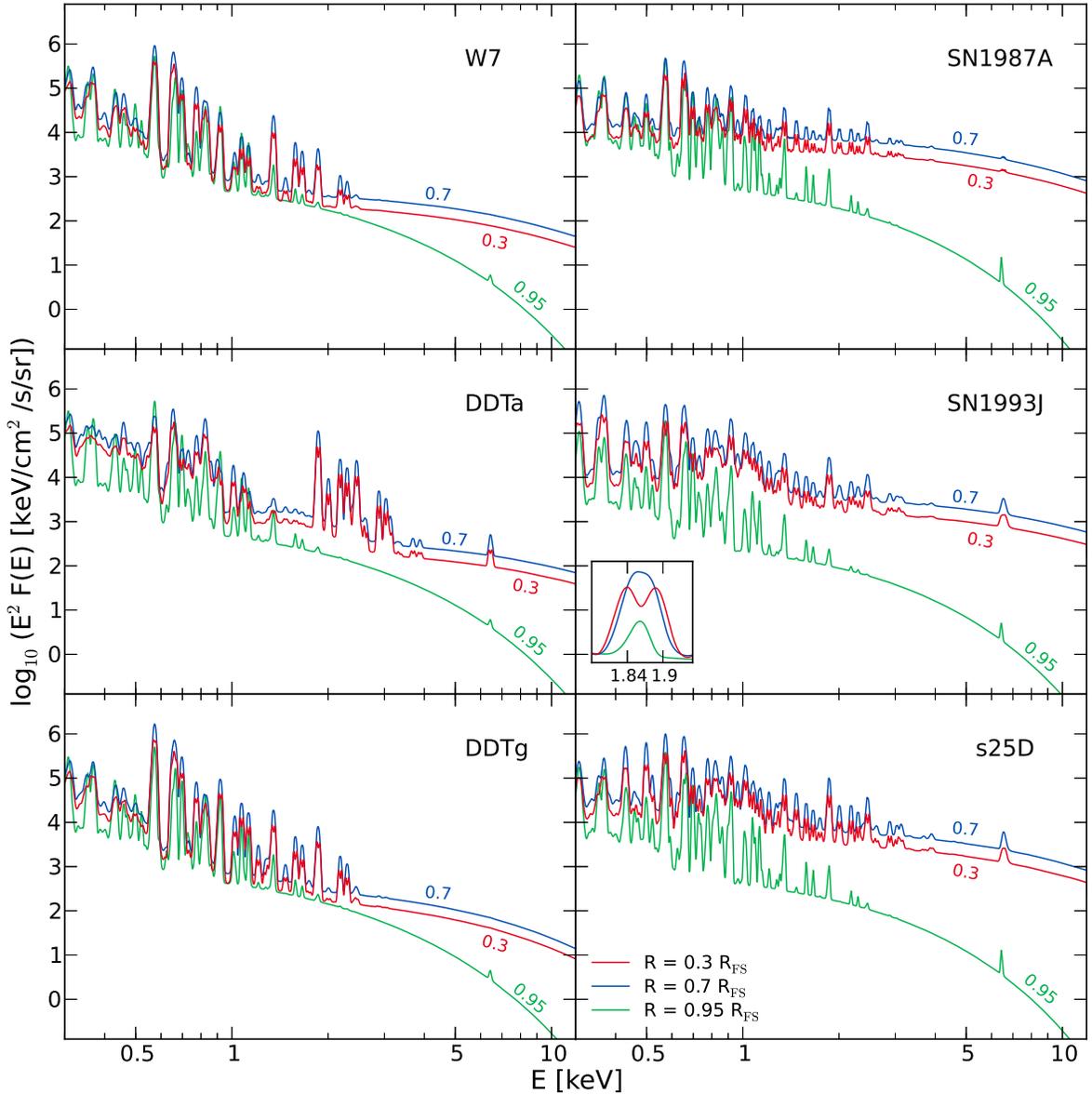}
\caption{
Line-of-sight emission along three paths with projected distances from the remnant center as indicated in the bottom right panel.
The widths of the extraction rings are all set 
to 10\% of $R_\mrm{FS}$, the spectra are normalized by the solid angle of the rings and convolved with a Gaussian with 
FWHM of 7~eV.
The inset in the panel for the SN1993J model compares the profiles of the Si-K line centered at $\sim 1.86$\,keV among the three regions, with the flux level normalized to each other at the respective line base.
}
\label{spec_los}
\end{figure*}

In Figure~\ref{spec_los} we show the LOS spectra taken at three different positions as noted.
The spectrum extracted close to the outer boundary of the SNR is mainly composed by emission from the shocked ambient gas in the CD-FS volume since the LOS does not cut through any of the shocked ejecta shells. As the LOS is moved inward, contributions from the shocked ejecta start to mix in, so the strong lines from, e.g., Si-K and S-K begin to appear in the spectra for the Type Ia models. 
Due to projection effects, the brightness generally increases as the LOS moves from the outermost boundary close to the FS (e.g. at $R = 0.95 \times R_\mrm{FS}$ shown in Figure~\ref{spec_los}) toward the inner positions.

In the inset shown in the panel for the SN1993J model, variation of the Si-K line profile with the LOS radial position is shown. 
The observed variation is consistent with a spherically expanding remnant
for which the Doppler shift of lines is symmetric,
and is the strongest when the LOS passes closest to the center of the remnant.
With the assumed instrumental spectral resolution of $E_\mrm{FWHM} = 7$~eV applied here, we see that a
clear splitting of lines can be resolved for projections through the inner regions of the SNR. 
We suggest that the ability to model this spitting will be useful for future spatially resolved spectroscopic studies of young SNRs such as Tycho. This Doppler broadening will probe the dynamics of the expanding plasma inside the remnant and provide information on
the explosion energetics and surrounding environment. 
In cases where proper motion measurements are also available, the distance to an SNR can be constrained as well 
\citep[see e.g.,][ for such an attempt with 
Tycho using \textit{Suzaku} and \textit{Chandra} data]{Hayato2010}.     

\subsection{Effect of Ambient Medium}
The environment of an SNR is a critical factor in determining the X-ray emission properties. The ambient density for example will influence the location of the RS in the ejecta at a given age and, depending on the progenitor and explosive nucleosynthesis models, the chemical composition of the shocked ejecta can vary drastically as the RS moves through the various ejecta layers.

\begin{figure}
\centering
\includegraphics[width=8cm]{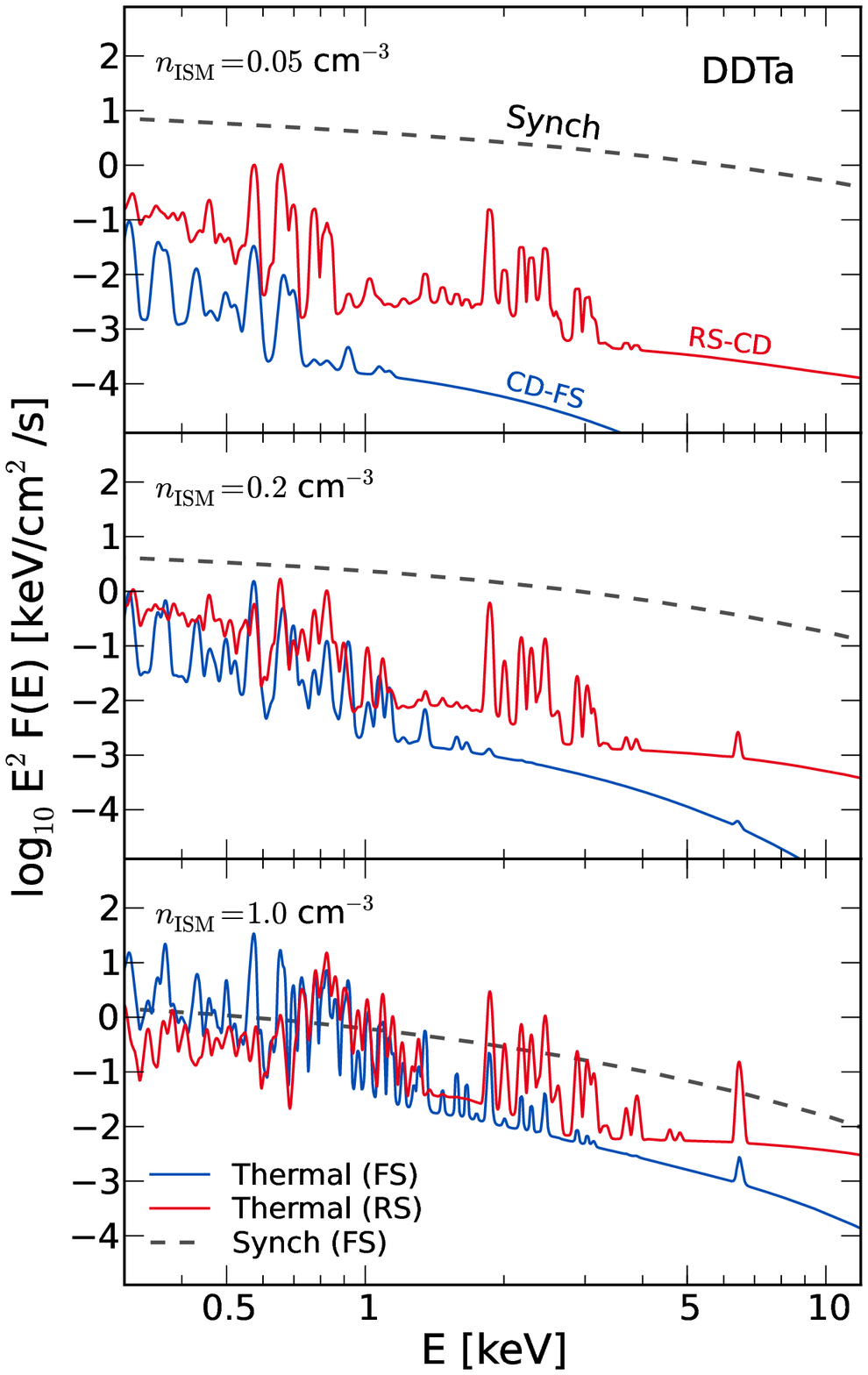} 
\includegraphics[width=8cm]{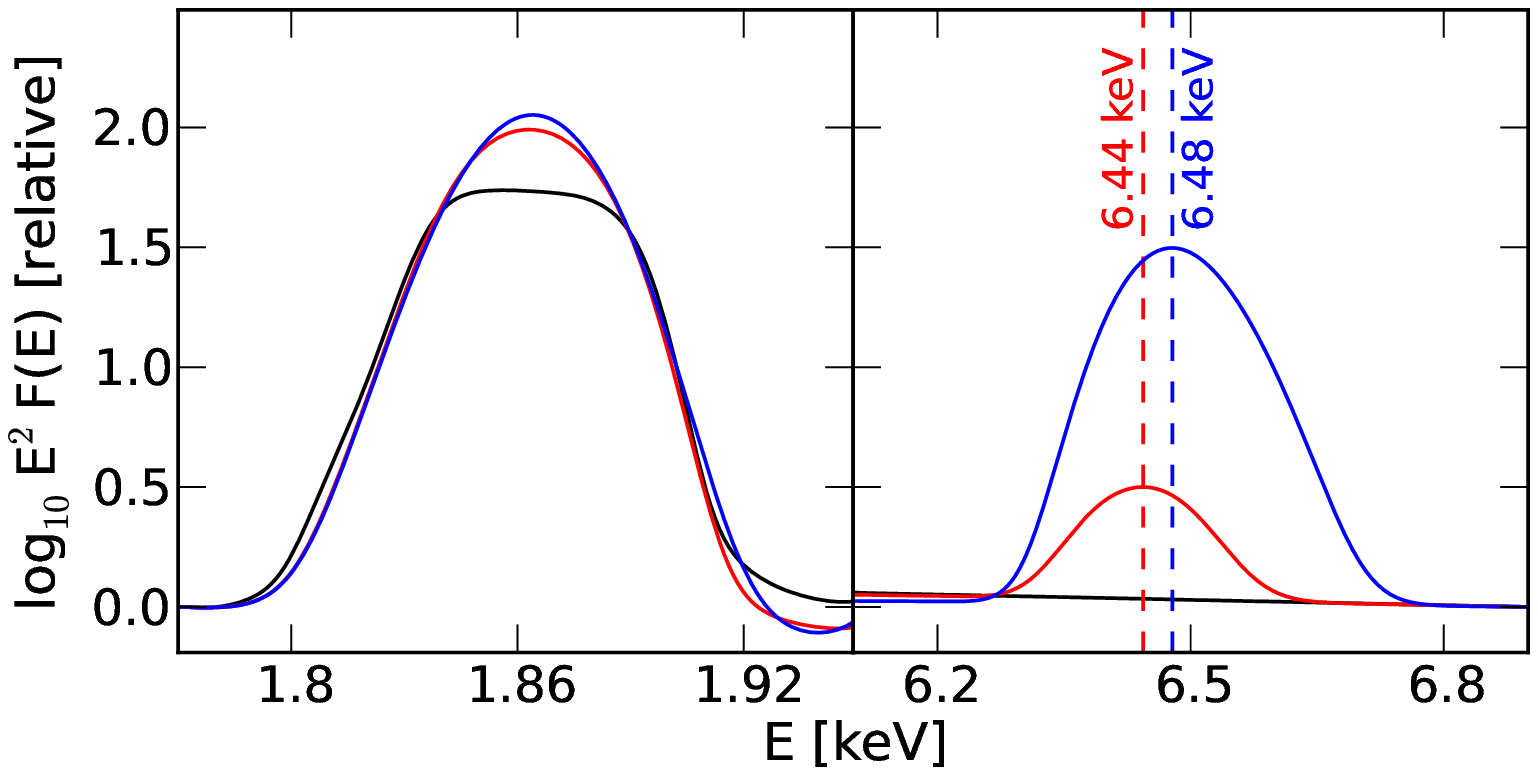}
\caption{
Top panel: Same as Figure~\ref{spec_integ} but for the Type Ia DDTa ejecta models with three different ambient medium densities: 
$\nISM=0.05$\,\pcc\ (top panel), $\nISM=0.2$\,\pcc\ (middle panel), and $\nISM=1.0$\,\pcc\ (bottom panel).
See text for a fuller description of the parameters used.
Bottom panel: Detailed view to compare the Si-K and Fe-K lines from the ejecta for the three models: 
black ($\nISM=0.05$\,\pcc), red ($\nISM=0.2$\,\pcc), and blue ($\nISM=1.0$\,\pcc). 
The flux levels are normalized to unity at the 
respective line base. The vertical dashed lines and accompanying numerics 
(red is 6.44\,keV and blue is 6.48\,keV) show the approximate centroids 
of the Fe-K lines. 
(A color version of this figure is available in the online journal.)
}
\label{spec_integ_den}
\end{figure}

In Figure~\ref{spec_integ_den}, we show volume-integrated 
X-ray spectra using the Type Ia DDTa ejecta model.
We add one low-density ($n_\mrm{ISM} = 0.05$~cm$^{-3}$) and 
one high-density ($n_\mrm{ISM} = 1.0$~cm$^{-3}$) case 
to the $n_\mrm{ISM} = 0.2$~cm$^{-3}$ example shown thus far.
The remnant expansion rate scales inversely with density and for these cases the FS speed is 6300, 4500, and 3000\,\kmps, respectively at 500~yr.
For the low-density case, not only does the normalization of the thermal spectrum from the shocked ISM scale as $n_\mrm{ISM}^2$, the faster adiabatic expansion of the ejecta also weakens the emission from the RS-CD volume. 
That the ejecta were expanding into a lower density medium also results in the creation of a weaker RS that propagates across 
less ejecta material in a given time ($R_\mrm{RS}/R_\mrm{FS} = 0.78$, 0.74 and 0.67 respectively), which results in weaker thermal emission. 
For example, the total amount of shocked ejecta Fe is 
$\sim 0.33$, 0.57 and 0.83\,$\Msun$ respectively for the three $\nISM$ values.
The faster expansion of the remnant in the low-density case also 
produces more Doppler broadening from the plasma bulk motion. 
In the bottom two panels of Figure~\ref{spec_integ_den} we show the Si-K and Fe-K lines from the ejecta for the three densities. The Si-K line for the $\nISM=0.05$\,\pcc\ case (left panel, black curve) shows a ``flat-top'' profile resulting from the combination of all line profiles through the remnant. For the other two cases with higher $\nISM$ values, as mentioned above, the expansion velocities of the shocked plasma in the observer's frame are not as high as the low-density case during the 500~yr of evolution, therefore this Doppler effect is less visible.

The number of lines also depends on the ambient density. In the high-density case, the heavy ions obtain higher ionization states and produce more lines.
This increase in ionization also produces a shift in the centroid of the Fe-K$\alpha$ line 
by $\sim 40$\,eV between $\nISM=0.2$\,\pcc\ and $\nISM=1$\,\pcc\ 
(right bottom panel in Figure~\ref{spec_integ_den}).
The mass-averaged integer charge states of the shocked ejecta Fe, by increasing density, are +5, +7 and +11.
   
Relatedly, we know that some special young CC SNRs like Cas A show strong Fe-K line emission that is detected from their outer regions; some parts appear to have an ``overturning'' of the nuclear burning layers where Fe-line emitting gas can be found exterior to the lighter elements like Si \citep{HRBS2000}. Some of these X-ray emitting ejecta also appear to expand at higher velocities relative to the optical main shell \citep[e.g.,][]{Delaney2010, Milisavljevic2013}. In fact, for Cas A, the total amount of shocked ejecta Fe is estimated to be 
$0.09 - 0.13$ $M_\odot$ \citep[e.g.,][]{Hwang2012}, indicating that almost all ejecta Fe has been shocked with very little left interior to the RS,
despite the young age of the remnant. 
We investigated the possibility of reproducing such a large amount of shocked Fe using our Type IIb SN1993J 
model tuned with Cas A parameters but ignoring any ``overturning'' of Fe and Si layers.
The amount of shocked Fe in our model is found to be $\sim 1.8 \times 10^{-3}\,\Msun$, almost two orders-of-magnitude lower than the observed mass and well below the amount needed to produce a strong 
Fe-K line as seen in Figure~\ref{spec_integ}. 

The reason is that, at 500~yr, the RS is still well separated from the Fe-rich core of the ejecta,
with $R_\mrm{RS}/R_\mrm{FS} = 0.78$ while the outer boundary of the Fe-rich region is at $\sim 0.35 \times R_\mrm{FS}$.
Increasing the pre-SN wind density by increasing $dM/dt$ by a factor of five causes the RS to reach deeper into the ejecta core but this is still insufficient to push the RS into the Fe-rich region. For this case, the shocked Fe mass is $\sim 3.1 \times 10^{-3}\,\Msun$ and the expansion speed of the shocked Fe in the ejecta is $\sim 2500$\,\kmps, too low compared to observation.

Since our spherically symmetric models cannot reproduce such a strong Fe-K line within a timescale of a few 100~yr, even if we invoke a high-density wind in the CSM, it confirms that other mechanisms are at work. Asymmetric effects and/or hydrodynamic instabilities have been proposed by a number of groups to bring heavy elements from the core to the exterior of the ejecta at high velocities during the SN explosion. This has been thoroughly investigated by a few groups recently through three-dimensional hydrodynamic simulations with explosive nucleosynthesis \citep[e.g.,][and references therein]{Ono2013}. 
However, it is important to note that Cas A may not be typical of CCSN remnants.
For example, some SNRs such as 1E 0102.2-7219 in the Small Magellanic Cloud are found to retain the hydrostatic 
onion structure of their progenitor stars without strong evidence of spatial inversion of the nuclear burning layers. The X-ray spectrum of 1E 0102.2-7219 shows little shocked Fe and the ejecta material has a relatively slow expansion velocity \citep[e.g.,][]{Flanagan2004}.

\subsection{Effect of Alfv\'{e}n Wave Damping in Shock Precursor}
\begin{figure}
\centering
\includegraphics[width=7.8cm]{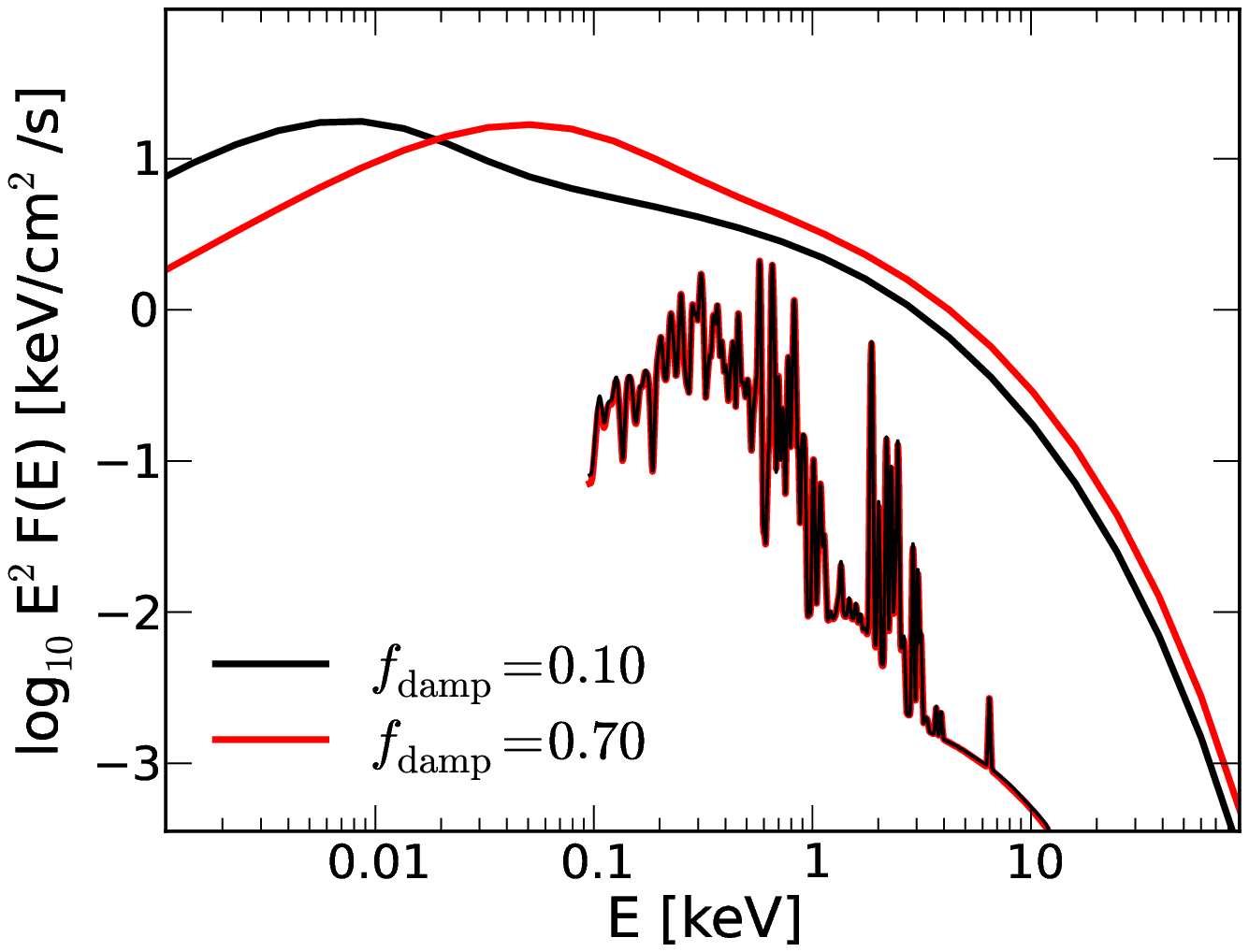}
\includegraphics[width=8cm]{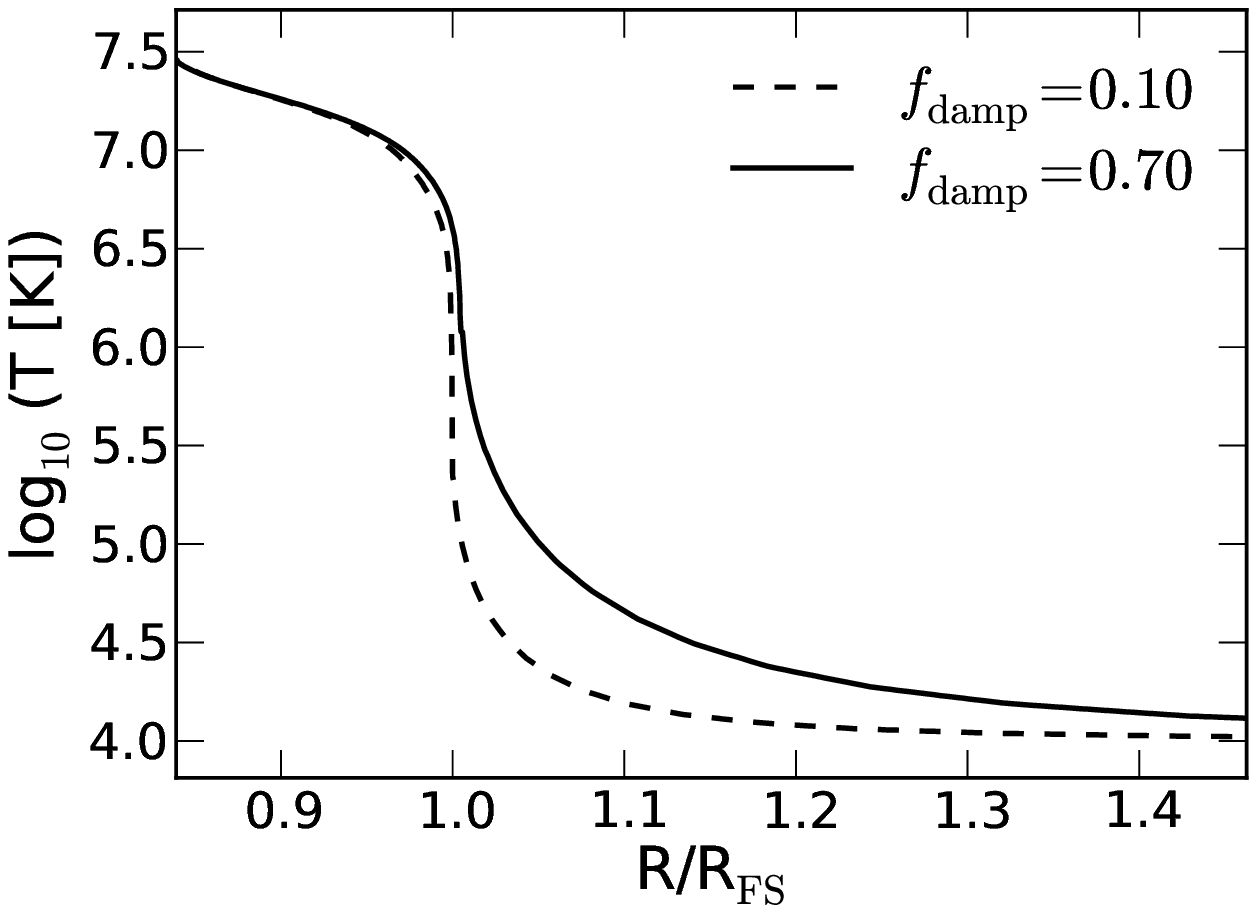}
\caption{
Top panel: The effect of Alfv\'{e}n wave damping rate in the FS precursor on the synchrotron (smooth curves) and thermal spectra. 
The thermal spectrum here is integrated over the whole 
remnant and the non-thermal spectrum is integrated over the CD-FS region.
The Type Ia DDTa ejecta model is used here. The results for a case with moderate damping (black lines) and a case with fast damping (red lines) are compared. No important influence on the total thermal spectrum is found. Bottom panel: Precursor and post-shock (CD-FS) electron temperature profile from the fast-damping case (solid) and moderate-damping case (dashed). The pre-heated temperatures of the plasma immediately in front of the subshock from the two models at 500~yr are about 50 and 320~eV respectively.
}
\label{spec_damp}
\end{figure}

Another factor that can be important for thermal X-ray emission when DSA is efficient is the  pre-heating of electrons in the CR-precursor.
Damping of CR-driven magnetic waves can deposit thermal energy to the precursor material and heat up the electrons and ions before they cross the dissipative subshock layer. Recent observations of fast Balmer-dominated shocks at non-radiative SNRs have suggested that fast equilibration of electrons and ions 
can heat electrons to $\sim 300$\,eV in front of the shock, independent of the shock speed \citep[e.g.,][]{GLR2007, Ghavamian2013}. 
We examine this effect by comparing results with the wave-damping parameter $\fDamp=0.1$ and $0.7$ in the DDTa model in 
Figure~\ref{spec_damp}.
The pre-heated plasma temperature immediately in front of the subshock for the $\fDamp=0.1$ and $0.7$ cases at 500~yr are about 50~eV and 320~eV respectively.  We found that although the pre-shock electron and ion temperatures are raised by more than a factor of six, the influence on the thermal X-ray spectrum is insignificant. 

A more significant effect, however, occurs for the non-thermal synchrotron emission.
The damped waves suppress the amplified magnetic field strength in the precursor (by a factor of two in this case). 
This increases the maximum momentum of the accelerated electrons, which is limited by synchrotron losses after 
$\sim 100$~yr, and results in a higher synchrotron cutoff energy (top panel).
On the other hand, the dynamics of the SNR such as the FS velocity is not affected as much, so that the shocked proton and ion temperatures stay mostly unaffected. The total compression ratio is decreased slightly from 8.5 to 7.9, producing only a small change to the 
post-shock densities. 
The electron temperature immediately behind the shock (bottom panel) is increased in the case of the high damping rate, but the subsequent equilibration with protons and ions quickly brings it back to values very similar to the $\fDamp=0.1$ model since the shocked proton and ion temperatures are not  strongly affected by the enhanced damping. Overall, only an insignificant change to the collisional ionization rate in the shocked plasma results.

\subsection{Effect of DSA Efficiency}

\begin{figure}
\centering
\includegraphics[width=8.8cm]{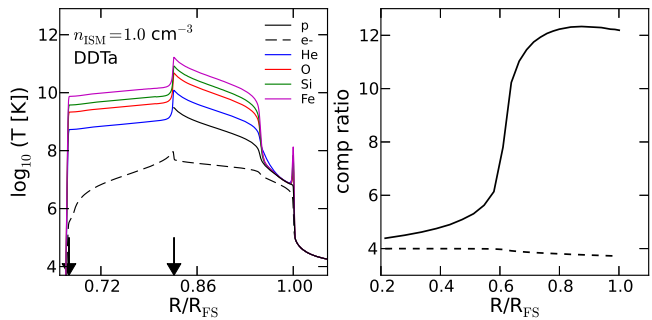}
\caption{
Illustration of the effect of nonlinear DSA at the FS on temperature evolution. The left panel shows the temperature profile at 500~yr for the DDTa ejecta model with~$n_\mrm{ISM} = 1.0$~cm$^{-3}$ (corresponds to the lower panel of Figure~\ref{spec_integ_den}). The right panel shows the evolution of the FS compression ratios $R_\mrm{tot}$ (total, solid line) and $R_\mrm{sub}$ (subshock, dashed line) as a function of FS radius in units of the FS radius at 500~yr. 
A rapid transition to a highly modified shock when $R_\mrm{FS}$ was about 60\% of the final radius is observed, corresponding to the point of sudden drop of $T_p$ and $T_i$ at around $R =  0.95 \times R_\mrm{FS}$ in the temperature profile \citep[see][]{BE99}. 
}
\label{temp_FS_eff}
\end{figure}

Highly efficient DSA at either the FS or RS can drastically decrease the shocked temperatures of protons and ions and also increase the shocked plasma density. 
These changes reduce the equilibration timescale $\Teq$ in a \NL\ fashion and result in lower final temperatures for the heavy ions.
As a result, narrow line profiles with suppressed thermal broadening can be expected in cases where efficient DSA takes place, and vice versa. 
On the other hand, the electron temperature will rise more quickly with efficient DSA which can have a 
direct influence on the post-shock ionization rates. 
As an example, Figure~\ref{temp_FS_eff} shows the temperature profile of a case where highly efficient DSA occurs at the FS. In this example, the instantaneous DSA efficiency reaches $\sim 90$\% at 500~yr and the FS is highly modified with extremely fast temperature equilibration.
Under appropriate conditions with an extremely efficient DSA, the temperature immediately behind the shock may drop further to as low as $\sim 10^6$~K. In this situation, the ionization rates of the freshly shocked heavy elements are expected to be hampered, despite a higher shocked gas density and faster collisional heating of the electrons.

\begin{figure}
\centering
\includegraphics[width=8.5cm]{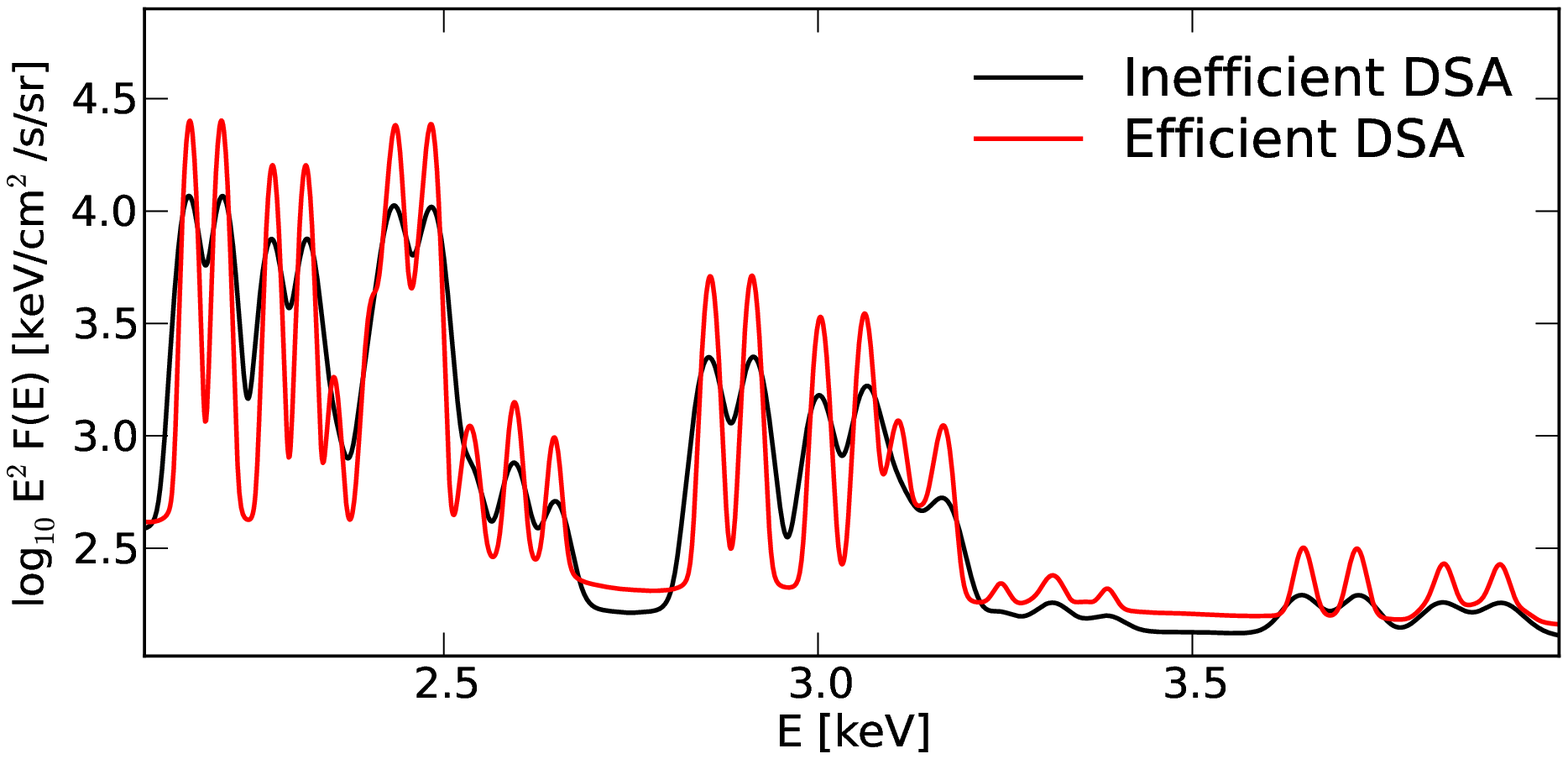}
\includegraphics[width=8.5cm]{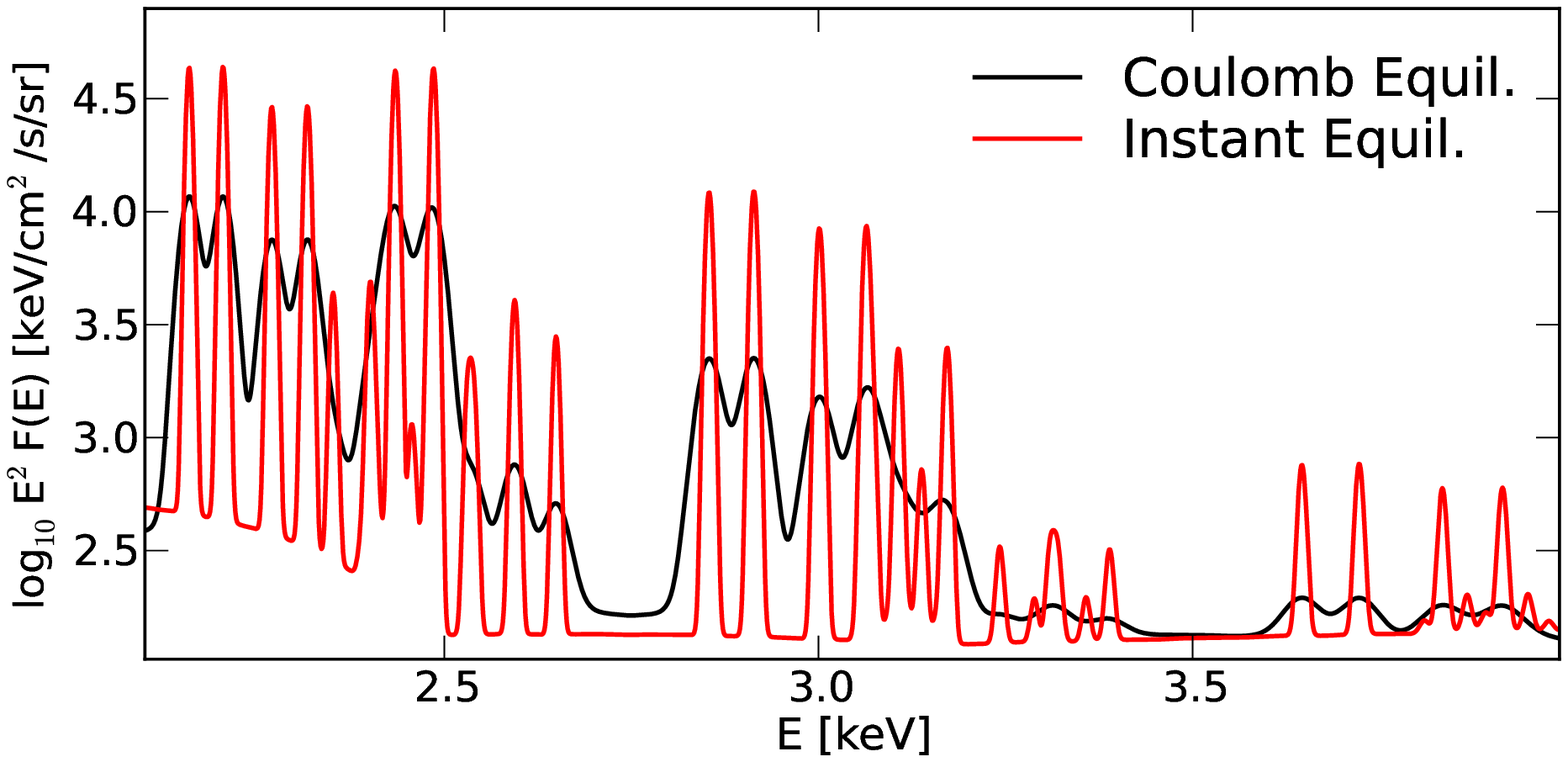}
\caption{
%
Illustration of the effect of DSA efficiency at the RS (top panel) and equilibration model (bottom panel) on the 
X-ray spectrum from the ejecta at 500~yr in the energy range of $2.1 - 4.0$~keV. Here the Type Ia DDTa ejecta model is used again. 
The spectra shown are obtained with LOS projection and are extracted from a ring of radius $R = 0.3 \times R_\mrm{FS}$ and width $\Delta R = 0.1 \times R_\mrm{FS}$. Lines originating from the fast-expanding plasma along the LOS are split into two by Doppler shifts, as expected.
In both panels, the black lines correspond 
to Coulomb equilibration and $\EffDSArs<0.01$, while the red lines use $\EffDSArs = 0.7$ (top panel) and instantaneous equilibration (bottom panel).
}
\label{spec_effects}
\end{figure}

While the existence of efficient DSA at the RS is controversial, there is evidence for DSA from X-ray observations of some young SNRs, such as the western part of Cas A \citep[e.g.,][]{HV2008}. Therefore, it is important to consider the effects of particle acceleration on X-ray emission from ejecta. In Figure~\ref{spec_effects} we compare the 
thermal X-ray spectrum at 500~yr from the shocked ejecta for the DDTa ejecta model in the energy range $2.1 - 4.0$~keV. The black curves in Figure~\ref{spec_effects} show the results with negligible DSA at the RS, while the red curves show the efficient case with $\EffDSArs = 0.7$ throughout the simulation.
The upstream $B$-field is fixed at $B_0 = 0.1$\,\muG\ to imitate MFA from efficient CR acceleration at the RS (see Section~\ref{section:dsa} for details). 
%
It is clear that the efficient conversion of the shock kinetic energy into non-thermal particles at the RS decreases the shocked ion temperatures and results in narrow line profiles.
Because of this, some weak satellite lines previously blended into the dominant lines can now be more clearly resolved. In principle, these lines could be observed by 
future X-ray telescopes with a spectral resolution at the eV-level, providing further important information on the plasma conditions in young SNRs.  

\subsection{Effect of Temperature Equilibration Model}

Thus far we have only presented examples with temperature equilibration among ions and electrons through slow Coulomb interactions, however, recent observations of young SNRs suggest that post-shock equilibration may be much faster than Coulomb collisions can produce \citep[e.g., see review by][]{Rakowski2005}. The latest analysis of spatially resolved Fe K$\alpha$ and Fe K$\beta$ line emission in the shocked ejecta of Tycho SNR by \citet{Yamaguchi2014} using the \textit{Suzaku} satellite has also revealed an electron temperature more than 100 times higher than that expected from pure Coulomb collisions, in regions not so far downstream from the RS. One possibility for this rapid heating is a fast post-shock temperature equilibration mediated by collisionless wave-particle interactions.
Here we consider a simple model for the extreme case in which a full temperature equilibration among all particle species is achieved instantaneously behind the shock. In the bottom panel of Figure~\ref{spec_effects}, for the ejecta emission with $\EffDSArs< 0.01$, we compare X-ray spectra with Coulomb equilibration (black curve) and instantaneous equilibration (red curve) using the Type Ia DDTa model.

In the energy range shown, narrow lines are predicted with instantaneous equilibration since heavy ions like S, Ar, and Ca producing these lines are cooled rapidly by equilibrating with lighter elements that have lower initial shocked temperatures. 
The difference between this effect and that stemming from efficient DSA discussed above is that, with DSA, the shocked temperatures of all ion species are lowered by a common factor through the energy loss to the non-thermal particle population, while in this case the net change from the shocked temperature, whether it is positive or negative,
can differ among elements and depends very much on the chemical composition of the 
equilibrating gas.
As a result, different lines can respond to these two effects differently. 
In reality, DSA efficiency and temperature equilibration may be coupled. 
However, by comparing well-resolved profiles of lines from different elements over a broad energy range, even in the case that both of these effects turn out to be important, it may be still possible to disentangle them from each other. 

Future X-ray observations by next-generation spectrometers such as SXS onboard \textit{Astro-H} will provide high-resolution spectral data in a broad energy band that can directly constrain equilibration models, or even discern efficient DSA at the RS without requiring information from any non-thermal emission, through detailed comparison with our broadband simulation models.   

\subsection{Time Evolution of X-ray Spectra}  

\begin{figure*}
\centering
\includegraphics[width=15cm]{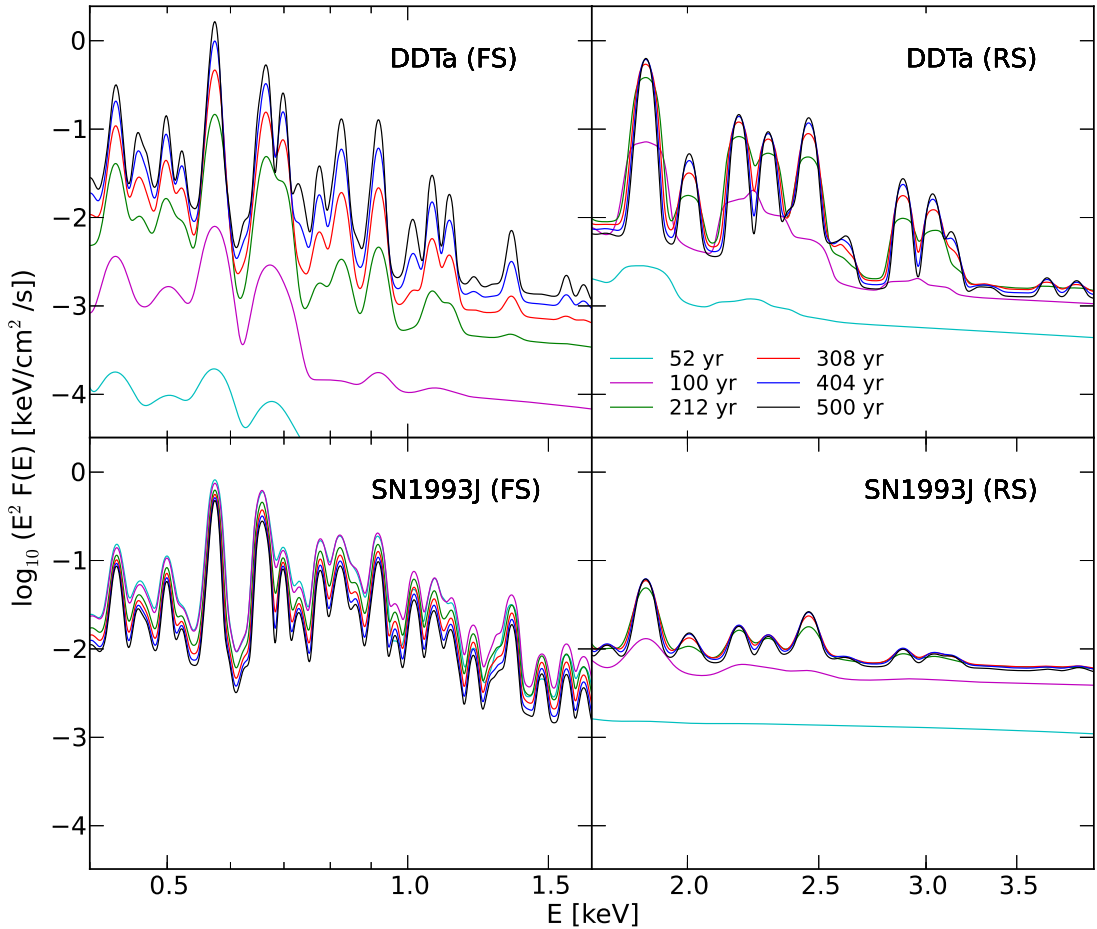}
\caption{
Time evolution of volume-integrated thermal X-ray spectra 
calculated at the ages shown in the top right panel.
The left and right panels show the spectra from the CD-FS region in the energy range of $0.4 - 1.7$~keV and from the RS-CD region in the energy range of $1.7 - 4.0$~keV, respectively.
(A color version of this figure is available in the online journal.)
}
\label{spec_evo_integ}
\end{figure*}  
%
Finally, we consider the rapid time evolution of the X-ray emission during the first few hundred years of the SNR lifetime.
When future high-resolution spectrometers observe ejecta-dominated SNRs of different ages, these models may help determine the evolutionary history of a given remnant.

Figure~\ref{spec_evo_integ} shows X-ray spectra from the RS-CD and CD-FS volumes for various ages from about 50~yr to 500~yr using a Type Ia model (top panels) and a CCSN model (bottom panels). The general trend of spectral evolution can be summarized as follows: in the first few decades, the shocked gas possesses a high temperature and high expansion velocity, causing severe thermal broadening and Doppler shifting of lines.
The combination of these broadenings can result in multiple lines blended into a single broad spectral feature, that is most obvious in the early ejecta spectrum for the DDTa model. The lines are also generally found mostly at lower energy since the number fraction of high ionization states of heavy elements responsible for the higher energy lines are still low. 

As time passes, the shock velocity drops and the ion temperature decreases, line profiles start to narrow, 
and individual lines previously blended are now better resolved. Lines from higher ionization states also begin to show up above the continuum. 
The increase in the overall flux level begins to slow since adiabatic dilution of the shocked gas starts to become important. This effect can compensate for the increase in new ejecta and ISM/CSM material swept up by the shocks which is also slowing with time.

An obvious difference in the evolutionary behavior of the shocked ISM/CSM material is that the thermal emission from the Type Ia model steadily increases while the CCSN emission gradually decreases with time.
In the Type Ia model, the FS is moving into a uniform low-density ISM medium, while the CCSN model has an $r^{-2}$ density gradient in the pre-SN RSG wind. 
As the FS propagates into the isotropic wind, the density upstream of the shock decreases as does the rate at which new CSM is swept up by the shock.
Adiabatic expansion can thus result in a flux level that is nearly independent of time, at least for remnants younger 
than $\sim 200$\, yr. This may be an important piece of information for discerning the properties of the ambient environment of X-ray bright young SNRs, particularly if long-term consecutive monitoring programs are carried out by current and future X-ray telescopes.
Of course this requires that the X-ray synchrotron flux is low enough such that the thermal spectrum can be measured from the CD-FS region, possibly from part of the azimuthal rim where DSA is inefficient. 
 
\section{Conclusions}
\label{conclusion}
We have presented a generalization of our spherically symmetric \crhydro\ code that includes thermal
X-ray emission at the RS, as well as the FS, in an evolving, 
shell-type SNR undergoing efficient DSA. 
While not attempting a detailed model of any 
particular SNR, we have investigated the effects of
various crucial physical processes and environmental factors that can alter the characteristics of the X-ray spectrum of young SNRs.
Our major results include: 

\begin{enumerate}
\item The \crhydro\ code now includes,
throughout the interaction volume between the RS and FS, 
the physics network of SNR hydrodynamics, \NL\ DSA, NEI, heavy-ion temperature equilibration, thermal X-ray line emission, and broadband \NT\ emission from various emission processes. \\

\item 
We incorporate the detailed ejecta chemical composition using current SN explosive nucleosynthesis simulations for both Type Ia thermonuclear SNe and CCSNe, trace the shock dynamics through these ejecta, and follow the ionization of each heavy element species using a fully time-dependent and spatially resolved method. \\

\item 
We have demonstrated the capability of the 
code to predict high-resolution thermal X-ray line spectra consistently with the broadband
non-thermal emission for different progenitor types and ambient environments. These predictions can be readily employed to model spectral data from current and next-generation X-ray spectrometers to extract important information about the properties and origin of an SNR. \\

\item
We vary the ambient gas density and show how this  
can directly change the dynamics of the expanding shocked ejecta, altering the ionization state of X-ray line-emitting elements, as well as the Doppler broadened line profiles as a function of time. \\

\item
We clearly show that a one-dimensional model cannot produce the strong Fe-K line observed from, e.g., Cas A, within $\sim 500$~yr even in a dense RSG wind, indicating that three-dimensional effects are indispensable for mixing the Fe-rich ejecta core of CCSNe up to the outer layers in order for it to be fully shocked by the RS. \\

\item
We vary the amount of wave damping in the CR precursor and 
show that efficient damping can increase the pre-shock temperatures 
of electrons and ions to $> 300$~eV, but this has a 
minimal effect on the final X-ray thermal spectrum. \\

\item
We include the effects of efficient DSA at the FS and RS and show how this process can lower the shocked ion temperatures and increase the shocked density. This shortens the inter-species equilibration timescale and sharpens the X-ray line profiles. \\

\item
We show how the temperature equilibration rate influences the X-ray spectrum and discuss rapid equilibration in light of recent observations that suggest this may be taking place in some young SNRs. \\

\item
We show how the X-ray emission as a function of energy depends on the chemical composition in the ejecta material, how this is coupled with the temperature equilibration rate, and how this varies as the RS enters different ejecta
layers. \\

\item
We calculate the time evolutionary behavior of the thermal
X-ray spectra for remnants from  typical Type Ia SNe expanding in uniform ISM, and CCSNe 
expanding in a pre-SN RSG wind. We show that these two cases have very different behaviors  and propose that monitoring observations during the first $\sim 200$\,yr could  provide critical information on the SN environment.
\end{enumerate}

We believe the generalized
\crhydro\ model we present here is the most fully \SC\ treatment of the thermal and \NT\ emission of an evolving SNR undergoing efficient CR production currently available. While the spherical symmetry we assume is clearly an important restriction, we expect this model will be an effective tool for future broadband modeling of SNRs,
especially with the high-resolution X-ray spectroscopic measurements, and hard X-ray imaging observations, expected from \AstroH.

The next step in our code development 
is to couple the \NL\ DSA aspects of our model to three-dimensional hydrodynamic simulations of SNRs.
This will allow us to investigate the origin of different morphological and broadband spectral features observed in the
SNR population. 
Another long-term goal is to estimate the contributions of various SN types to the total energy budget and spectrum of Galactic CRs.
Finally, we note that in principle, it should be possible to combine our model with three-dimensional
SN explosion simulations allowing a seamless description of the evolution from the progenitor star into the remnant phase.

\acknowledgements
We thank C. Badenes, M. Hashimoto, A. Heger, T. Nozawa, M. Ono and T. Shigeyama for generously providing SN progenitor and nucleosynthesis data for the completion of this work. We also express special thanks to C. Badenes and the anonymous referees for their valuable comments and suggestions for the improvement of the manuscript. S.-H. L. acknowledges support from Grants-in-Aid for Foreign JSPS Fellow (No. 2503018). D.J.P. and P.O.S. acknowledge support from NASA contract NAS8-03060. D.C.E. acknowledges support from NASA grant NNX11AE03G. S.N. acknowledges support from the Japan Society for the Promotion of Science (Nos. 23340069, 24.02022, 25.03786 and 25610056).

\bibliographystyle{aa} 
\bibliography{reference}

\end{document}